\documentclass[12pt]{article}
\pdfoutput=1

\usepackage[bulletsep]{collref}
\usepackage{amssymb,graphicx}
\usepackage[intlimits]{amsmath}
\usepackage{bbm}
\usepackage{mathtools}
\usepackage{MnSymbol}

\usepackage{slashed}

\usepackage{hyperref}
\usepackage{url}

\makeatletter \@addtoreset{equation}{section} \makeatother

\makeatletter
\let\old@startsection=\@startsection
\let\oldl@section=\l@section
\renewcommand{\@startsection}[6]{\old@startsection{#1}{#2}{#3}{#4}{#5}{#6\mathversion{bold}}}
\renewcommand{\l@section}[2]{\oldl@section{\mathversion{bold}#1}{#2}}
\makeatother

\makeatletter
\let\old@makecaption=\@makecaption
\def\@makecaption{\small\old@makecaption}
\makeatother

\newcommand{\bea}{\begin{eqnarray}}
\newcommand{\eea}{\end{eqnarray}}
\def\be{\begin{equation}}
\def\ee{\end{equation}}
\newcommand{\bei}{\begin{itemize}}
\newcommand{\eei}{\end{itemize}}
\newcommand{\bee}{\begin{enumerate}}
\newcommand{\eee}{\end{enumerate}}

\def\am{{\rm am}}
\def\am0{{\rm am}_0}

\begin{document}


\begin{flushright}\footnotesize
\texttt{NORDITA-2017-15} \\
\texttt{UUITP-06/17}
\vspace{0.6cm}
\end{flushright}

\renewcommand{\thefootnote}{\fnsymbol{footnote}}
\setcounter{footnote}{0}

\begin{center}
{\Large\textbf{\mathversion{bold} Quantum String Test\\ of Nonconformal Holography}
\par}

\vspace{0.8cm}

\textrm{Xinyi~Chen-Lin, Daniel~Medina-Rincon and
Konstantin~Zarembo\footnote{Also at ITEP, Moscow, Russia}}
\vspace{4mm}

\textit{Nordita,  Stockholm University and KTH Royal Institute of Technology,
Roslagstullsbacken 23, SE-106 91 Stockholm, Sweden}\\
\textit{Department of Physics and Astronomy, Uppsala University\\
SE-751 08 Uppsala, Sweden}\\
\vspace{0.2cm}
\texttt{xinyic@nordita.org, d.r.medinarincon@nordita.org, zarembo@nordita.org}

\vspace{3mm}


\par\vspace{1cm}

\textbf{Abstract} \vspace{3mm}

\begin{minipage}{13cm}
We compute L\"uscher corrections to the effective string tension in the Pilch-Warner background, holographically dual to $\mathcal{N}=2^*$ supersymmetric Yang-Mills theory. The same quantity can be calculated directly from field theory by solving the localization matrix model at large-$N$. We find complete agreement between the field-theory predictions and explicit string-theory calculation at strong coupling.
\end{minipage}
\end{center}

\vspace{0.5cm}


\newpage
\setcounter{page}{1}
\renewcommand{\thefootnote}{\arabic{footnote}}
\setcounter{footnote}{0}

\section{Introduction}

Holographic duality acts most simply at strong coupling, in the regime where field-theory calculations are obviously difficult, and direct tests of holography are few beyond the most symmetric cases of $\mathcal{N}=4$ super-Yang-Mills (SYM) theory or ABJM model. These models are conformally invariant. Massive, non-conformal theories are much less explored in this respect. The $\mathcal{N}=2^*$ SYM, a close relative of $\mathcal{N}=4$ SYM where the adjoint hypermultiplet gets mass, is a lucky exception. This theory is simple enough to admit exact solution at strong coupling and at the same time has an explicitly known holographic dual \cite{Pilch:2000ue,Buchel:2000cn}.

On the field-theory side, supersymmetric localization computes the path integral of the $\mathcal{N}=2^*$ theory on $S^4$ without any approximations \cite{Pestun:2007rz}, resulting in a zero-dimensional matrix model. In order to access the holographic regime one needs to solve this model in the large-$N$ limit and then take the 't~Hooft coupling $\lambda =g_{\rm YM}^2N$ to be also large. The strong-coupling solution of the $\mathcal{N}=2^*$ matrix model is relatively simple \cite{Buchel:2013id}, and allows one to calculate the Wilson loop expectation value for any asymptotically large contour. The result is reproduced by the area law in the dual holographic geometry \cite{Buchel:2013id}. The free energy of the matrix model agrees with the supergravity action evaluated on the counterpart of the Pilch-Warner background with the $S^4$ boundary \cite{Bobev:2013cja}. These results are valid at strictly infinite coupling. The next order in the strong-coupling expansion of the localization matrix model was computed  in \cite{Chen:2014vka,Zarembo:2014ooa}. Our goal is to go beyond the leading order on the string side of the holographic duality.

Wilson loops in the $\mathcal{N}=2^*$ theory are defined as
\begin{equation}\label{WilsonLoop}
 W(C)=\left\langle \frac{1}{N}\,
 \mathop{\mathrm{tr}}{\rm P}\exp\left[
 \oint_C ds\left(iA_\mu \dot{x}^\mu +|\dot{x}|\Phi \right)
 \right]
 \right\rangle,
\end{equation}
where $\Phi $ is the scalar field from the vector multiplet. Their expectation values obey the perimeter law: 
\begin{equation}\label{perlaw}
 W(C)\stackrel{{\rm{M}}{\rm{L}}\gg 1}{=}\,{\rm e}\,^{T(\lambda ){\rm{M}}{\rm{L}}},
\end{equation}
for sufficiently large contours. Here $\rm{L}$ is the length of the closed path $C$ and $\rm{M}$ is the hypermultiplet mass. The coefficient of proportionality $T(\lambda )$ can be called effective string tension, since at strong coupling it is dictated by the area law in the dual geometry and takes on the standard AdS/CFT value $T=\sqrt{\lambda }/2\pi $. The strong-coupling solution of the localization matrix model is in agreement with this prediction \cite{Buchel:2013id}. The subleading order of the strong-coupling expansion has been also calculated on the matrix model side \cite{Chen:2014vka,Zarembo:2014ooa}:
\begin{equation}\label{T(lambda)}
 T(\lambda )=\frac{\sqrt{\lambda }}{2\pi }-\frac{1}{2}+\mathcal{O}\left(\frac{1}{\sqrt{\lambda }}\right).
\end{equation}
On the string-theory side of the duality the subleading term should come from quantum corrections in the string sigma-model, which we are going to analyze in this paper.

This is interesting for two reasons. Corrections in $1/\sqrt{\lambda }$  probe holography at the quantum level. String quantization in curved Ramond-Ramond backgrounds such as the Pilch-Warner solution is a highly non-trivial problem, not devoid of conceptual issues. Potential agreement with rigorous field-theory results is a strong consistency check on the formalism.

Another reason is a highly non-trivial phase structure of the localization matrix model which features infinitely many phase transitions that accumulate at strong coupling \cite{Russo:2013qaa,Russo:2013kea}. Holographic description of these phase transitions remains a mystery. The phase transitions occur due to irregularities in the eigenvalue density of the matrix model. The leading order of the strong-coupling expansion originates from the bulk of the eigenvalues density where irregularities are averaged over, while the subleading term in (\ref{T(lambda)}) is sensitive to the endpoint regime \cite{Chen:2014vka}, the locus from which the critical behaviour originates.

\section{The Pilch-Warner background}\label{intropwbackground}

Holography maps an expectation value of a Wilson loop to the partition function of a string with ends anchored to the contour on the boundary of the dual geometry \cite{Maldacena:1998im}:
\begin{equation}
 W(C)=\int_{C=\partial \Sigma }^{}\mathcal{D}X^M\,\,{\rm e}\,^{-S_{\rm string}[X]}.
\end{equation}
 The holographic dual of $\mathcal{N}=2^*$ SYM is the Pilch-Warner (PW) solution of type IIB supergravity \cite{Pilch:2000ue}. In this section we review the PW background. Our notations and conventions are summarized in appendix~\ref{conventionsandnotations}.

The Einstein-frame metric for the PW background is\footnote{In the notations of \cite{Pilch:2000ue,Buchel:2000cn,Pilch:2003jg}, $A=\rho ^6$. 
We also redefined $\theta \rightarrow \pi /2-\theta $ compared to these references.
From now on we set $\rm{M}=1$. The dependence on $\rm{M}$ can be easily recovered by dimensional analysis.} \cite{Pilch:2000ue,Pilch:2003jg}:
\begin{eqnarray}\label{Einsteinm}
 ds_E^2&=&\frac{\left(cX_1X_2\right)^{\frac{1}{4}}}{\sqrt{A}}\left[
 \frac{A}{c^2-1}\,dx^2+\frac{1}{A\left(c^2-1\right)^2}\,dc^2+\frac{1}{c}\,d\theta ^2
 +\frac{\cos^2\theta }{X_2}\,d\phi ^2\right.
\nonumber \\
&&\left.\vphantom{ +\frac{\cos^2\theta }{X_2}\,d\phi ^2}
+A\sin^2\theta \,d\Omega ^2
 \right],
\end{eqnarray}
where $c\in\left[ {1,\infty } \right)$ and $d\Omega ^2$ is the metric of the deformed three-sphere:
\begin{equation}\label{dOmega}
 d\Omega ^2=\frac{\sigma _1^2}{cX_2}+\frac{\sigma _2^2+\sigma _3^2}{X_1}\,.
\end{equation}
The one-forms $\sigma _i$ ($i=1,2,3$) satisfy 
\begin{equation}\label{Maurer-C}
 d\sigma_i=\epsilon _{ijk}\sigma_{j}\wedge\sigma_{k},
\end{equation}
and are defined in the $SU(2)$ group-manifold representation of $S^3$, as:
\begin{equation}
 \sigma _i=\frac{i}{2}\,\mathop{\mathrm{tr}}(g^{-1}\tau _idg),\qquad g\in SU(2), 
\end{equation}
 where $\tau _i$ are the Pauli matrices. The function $A$ is given by:
\begin{equation}\label{DefinitionofA}
 A=c-\frac{c^2-1}{2}\ln\frac{c+1}{c-1}\,,
\end{equation}
while $X_{1,2}$ are:
\begin{eqnarray}
 X_1&=&\sin^2\theta +cA\cos^2\theta ,
\nonumber \\
X_2&=&c\sin^2\theta +A\cos^2\theta .
\end{eqnarray}

The dilaton-axion is given by:
\begin{equation}
 \,{\rm e}\,^{-\Phi }-i C_{(0)} =\frac{1+\mathcal{B}}{1-\mathcal{B}}\,,\qquad 
 \mathcal{B}=\,{\rm e}\,^{2i\phi }\,\frac{\sqrt{cX_1}-\sqrt{X_2}}{\sqrt{cX_1}+\sqrt{X_2}} ,
\end{equation}
while the two-form potential $A_{(2)}=C_{(2)}+i B_{(2)}$ is defined as:
\begin{equation}\label{definitionA2}
{A_{(2)}} = {e^{i\phi }}\left( {{a_1}\ d\theta  \wedge {\sigma_1} + {a_2}\ {\sigma_2} \wedge {\sigma_3} + {a_3}\  {\sigma_1} \wedge d\phi } \right) ,
\end{equation}
with:
\begin{align}\label{aa1}
{a_1}\left( {c,\theta } \right) &= \frac{i}{c}{\left( {{c^2} - 1} \right)^{1/2}}{\mathop{\rm sin}\nolimits} \theta\ ,\\
\label{aa2}
{a_2}\left( {c,\theta } \right) &= i\frac{A}{{{X_1}}}{\left( {{c^2} - 1} \right)^{1/2}}{{\mathop{\rm sin}\nolimits} ^2}\theta\  {\mathop{\rm cos}\nolimits} \theta\ ,\\
\label{aa3}
{a_3}\left( {c,\theta } \right) &=  - \frac{1}{{{X_2}}}{\left( {{c^2} - 1} \right)^{1/2}}{{\mathop{\rm sin}\nolimits} ^2}\theta\  {\mathop{\rm cos}\nolimits} \theta\ ,
\end{align}
and the four-form potential $C_{(4)}$ is given by:
\begin{equation}
{C_{(4)}} = 4\omega {\kern 1pt} {\kern 1pt} d{x^0} \wedge d{x^1} \wedge d{x^2} \wedge d{x^3} ,
\end{equation}
where $\omega=\omega(c,\theta)$ is defined as:
\begin{equation}
\omega \left( {c,\theta } \right) = \frac{{A{\kern 1pt} {X_1}}}{{4{{\left( {{c^2} - 1} \right)}^2}}} .
\end{equation}

In terms of these potentials, the NS-NS three-form is given by $H=dB_{(2)}$, while the ``modified'' R-R field strengths are given by:
\begin{align}
{\tilde{F}_{(1)}} &= d{C_{(0)}} ,\\
{\tilde{F}_{(3)}} &= d{C_{(2)}} + {C_{(0)}}{\kern 1pt} d{B_{(2)}} ,\\
{\tilde{F}_{(5)}} &= d{C_{(4)}} + {C_{(2)}} \wedge d{B_{(2)}} = d{C_{(4)}} +  * d{C_{(4)}} ,
\end{align}
where ${\tilde{F}_{(5)}}$ satisfies $* {\tilde{F}_{(5)}} = {\tilde{F}_{(5)}}$.

\section{Setup}

Since the perimeter law (\ref{perlaw}) is universal, any sufficiently large contour can be used to calculate the effective string tension. The simplest choice is the straight infinite line regularized by a cutoff at length $\rm{L}\gg 1$. The minimal surface with this boundary is an infinite wall:
\begin{equation}\label{classs}
 x^1_{\rm cl}=\tau ,\qquad c_{\rm cl}=\sigma .
\end{equation}
This solution approximates the minimal surface  for any sufficiently big but finite contour on distance scales small compared to the contour's curvature. Eventually the true minimal surface turns around at some $c_0\sim \rm{L}\gg 1$ and goes back to the boundary.
As shown in \cite{Buchel:2013id}, the finite holographic extent of the minimal surface can be ignored in calculating the minimal area, which can thus be evaluated on the simple solution (\ref{classs}) upon imposing the large-distance cutoff $\rm{L}$. We will make the same assumption in calculating quantum corrections to the minimal area law, and will study quantum fluctuations of the string around the simple infinite-wall configuration. 

We also need to specify the position of the minimal surface on the deformed $S^5$. The $S^5$ part of the geometry is dual to scalars on the field-theory side, and the location of the string on $S^5$ is dictated by the scalar coupling of the Wilson loop (\ref{WilsonLoop}):
\begin{equation}\label{class-s5}
 \theta _{\rm cl}=0,\qquad  \phi _{\rm cl}=0,
\end{equation}
which completely specifies the string configuration, since  the three-sphere shrinks to a point at $\theta =0$.

The induced string-frame metric on the minimal surface, rescaled by a factor of $\,{\rm e}\,^{\Phi /2}|_{\rm cl}=1/\sqrt{\sigma}$ compared to the Einstein metric in (\ref{Einsteinm}),  is
\begin{equation}\label{wsmetric}
 ds^2_{\rm w.s.}=\frac{A}{\sigma ^2-1}\,d\tau ^2+\frac{1}{A(\sigma ^2-1)^2}\,d\sigma ^2,
\end{equation}
where now $A\equiv A(\sigma)$. The regularized sigma-model action evaluated on this solution equals to
\begin{equation}\label{leadingScl}
 S_{\rm reg}=\frac{\sqrt{\lambda }}{2\pi }
 \int_{{\rm reg}}^{}\frac{d\tau \,d\sigma }{(\sigma ^2-1)^{\frac{3}{2}}}
 =-\frac{\sqrt{\lambda }}{2\pi }\,{\rm{L}},
\end{equation}
where integration over $\tau $ and  $\sigma $ ranges from $-\rm{L}/2$ to $\rm{L}/2$ and from $1+\epsilon ^2/2$ to infinity, and the divergent $1/\epsilon $ term is subtracted by regularization. The area law in the PW geometry therefore agrees with the leading-order strong coupling result (\ref{perlaw}), (\ref{T(lambda)}) obtained from localization.

Our goal is to calculate holographically the $\mathcal{O}(\lambda ^0)$ term in the Wilson loop expectation value. 
The next order at strong coupling comes from two related but distinct sources. One is quantum fluctuations of the string and the other is the Fradkin-Tseytlin term in the classical string action, which is closely related to conformal anomaly cancellation and comes without a factor of $1/\alpha '\sim \sqrt{\lambda }$ \cite{Fradkin:1984pq,Fradkin:1985ys}. The Fradkin-Tseytlin term is usually ignored in holographic calculations of Wilson loops. This is justified for backgrounds with a constant dilaton,  for instance $AdS_5\times S^5$, where the Fradkin-Tseytlin term is purely topological. But  in the Pilch-Warner geometry the dilaton has a non-trivial profile and the Fradkin-Tseytlin term has to be taken into account.

It has been long recognized that string fluctuations play an important role in gauge-theory strings and are necessary, for example, to accurately describe the quark-anti-quark potential in QCD \cite{Luscher:2002qv}. The first quantum correction to the potential for the free bosonic string is the universal L\"uscher term \cite{Luscher:1980fr,Luscher:1980ac}. The free string can be quantized exactly and all higher-order fluctuation corrections can be explicitly calculated \cite{Alvarez:1981kc,Arvis:1983fp,Olesen:1985pv}. Holographic string, however, is not free, as it propagates in a complicated curved background, and one is bound to rely on perturbation theory. The first order, equivalent to the the L\"uscher term for the Nambu-Goto string,  involves expanding  the action of the string sigma-model around the minimal surface and integrating out the fluctuation modes in the one-loop approximation \cite{Greensite:1999jw,Forste:1999qn,Kinar:1999xu}. The
full-fledged formalism  for the background-field quantization of the string sigma-model in $AdS_5\times S^5$ was developed  in \cite{Drukker:2000ep} and has been successfully used to compute L\"uscher corrections to the static potential in $\mathcal{N}=4$ SYM \cite{Chu:2009qt,Forini:2010ek}. In that case the L\"uscher correction can actually be reproduced directly from field theory \cite{Gromov:2016rrp} using integrability of the AdS/CFT system \cite{Drukker:2012de,Correa:2012hh}.

The formalism of \cite{Drukker:2000ep}, originally developed for strings in $AdS_5\times S^5$, uses the Green-Schwarz string action expanded to second order in fermions, which is known for any supergravity background \cite{Cvetic:1999zs}. The semiclassical quantization of the Green-Schwarz superstring along the lines of  \cite{Drukker:2000ep} can thus be adapted to the PW geometry with minimal modifications.  Schematically, the embedding coordinates of the string are expanded near the classical solution: $X^\mu =X^\mu _{\rm cl}+\xi ^\mu $ to the quadratic order: $S[X]=S_{\rm cl}+\left\langle \xi ,\mathcal{K}\xi \right\rangle$. Gaussian integration over $\xi ^\mu $ then yields:
\begin{equation}\label{generalrationdets}
 W(C)=\,{\rm e}\,^{-S_{\rm cl}}\,\frac{\det^{\frac{1}{2}}\mathcal{K}_F}{\det^{\frac{1}{2}}\mathcal{K}_B},
\end{equation}
where $\mathcal{K}_B$ and $\mathcal{K}_F$ are quadratic forms for bosonic and fermionic fluctuations of the string, and $S_{\rm cl}$ is the string action evaluated on the classical solution. As discussed above, $S_{\rm cl}$ includes the Fradkin-Tseytlin term which is of the same order in $1/\sqrt{\lambda }$ as the one-loop partition function. 

In the next three sections we calculate the Fradkin-Tseytlin contribution to the classical action, derive the explicit form of the operators $\mathcal{K}_{B,F}$ and then compute the ratio of determinants that appears in (\ref{generalrationdets}).

\section{Fradkin-Tseytlin term}

The bosonic part of the sigma-model Lagrangian is
\begin{align}\label{cveticbosons}
{L_{B}} =  \frac{1}{2}\sqrt h {h^{ij}}{\partial _i}{X^\mu }{\partial _j}{X^\nu }{G_{\mu \nu }} + \frac{i}{2}{\epsilon ^{ij}}{\partial _i}{X^\mu }{\partial _j}{X^\nu }{B_{\mu \nu }} ,
\end{align}
where $G_{\mu\nu}$ denotes the background metric in the string frame and $B_{\mu \nu }$ is the $B$-field. We fix the diffeomorphism gauge by identifying the internal metric $h_{ij}$ with the induced metric on the classical solution (\ref{wsmetric}).

The Fradkin-Tseytlin term couples the two-dimensional curvature to the dilaton \cite{Fradkin:1984pq,Fradkin:1985ys}:
\begin{equation}
L_{\rm FT} =\frac{1}{4\pi } \,\sqrt{h}\,R^{(2)}\Phi .
\end{equation}
The coefficient in front is fixed by the relationship between the string coupling and the dilaton expectation value: $g_{\rm str}=\,{\rm e}\,^{\left\langle \Phi \right\rangle}$. The genus-$g$ string amplitude is then accompanied by the correct power of the coupling: $g_{\rm str}^{2-2g}$, in virtue of the Gauss-Bonnet theorem.

The full bosonic action of the sigma-model is
\begin{equation}
 S_B=\int_{}^{}d^2\sigma \,\left(\frac{\sqrt{\lambda }}{2\pi }\,L_B
 +L_{\rm FT}\right),
\end{equation}
where the sigma-model part of the classical action is calculated in (\ref{leadingScl}). We proceed with evaluating the Fradkin-Tseytlin term.

The curvature of the induced world-sheet metric (\ref{wsmetric}) is equal to
\begin{equation}
 \sqrt{h}R^{(2)}=2\frac{d}{d\sigma }(\sigma ^2-1)^{-\frac{1}{2}},
\end{equation}
which is a total derivative as it should be. For the dilaton evaluated on the classical solution, we have:
\begin{equation}
 \Phi |_{\rm cl} =-\ln\sigma .
\end{equation}
Integration  by parts gives
\begin{equation}
 S_{\rm FT}=\frac{2{\rm{L}}}{4\pi }\int_{1}^{\infty }\frac{d\sigma }{\sigma \sqrt{\sigma ^2-1}}=\frac{{\rm{L}}}{4}\,.
\end{equation}
Combining the result with (\ref{leadingScl}), we get:
\begin{equation}\label{sclass}
 S_{\rm cl}=\left(-\frac{\sqrt{\lambda }}{2\pi }+\frac{1}{4}\right){\rm{L}}.
\end{equation}
The Fradkin-Tseytlin term thus gives half of the expected correction to the effective string tension at strong coupling, if one compares with the result (\ref{T(lambda)}) predicted from localization. The genuine quantum corrections should be responsible for the other half.

\section{Bosonic fluctuations}\label{bosonsPW}

The background metric can be simplified in the vicinity of the classical world-sheet, since we only need to expand it to the second order in deviations from the classical solution (\ref{classs}). For the conformal factor in the string frame we get:
\begin{equation}
 \,{\rm e}\,^{\frac{\Phi }{2}}\,\frac{\left(cX_1X_2\right)^{\frac{1}{4}}}{\sqrt{A}}
 =1+\frac{c^2-1}{2}\,\phi ^2+\frac{c-A}{2A}\,{\theta} ^2+\ldots 
\end{equation}

The deformed three-sphere shrinks to a point on the classical solution. Importantly, the coefficients of the two terms in (\ref{dOmega}) become equal on the locus (\ref{class-s5}), after which the metric becomes proportional to that of  the round sphere. Up to $\mathcal{O}({\theta} ^2)$ corrections,
\begin{equation}
 d\Omega ^2\simeq \frac{\sigma _i^2}{Ac}=\frac{d\mathbf{n}^2}{Ac}\,,
\end{equation}
where $\mathbf{n}$ is a unit four-vector. In the $SU(2)$ parametrization, $g=n^0+in^i\tau _i$. Introducing the Cartesian four-vector in the tangent space,
\begin{equation}\label{def-y}
 \mathbf{y}={\theta} \mathbf{n},
\end{equation}
we find that the $d{\theta} ^2$ and $d\Omega ^2$ terms in the Pilch-Warner metric combine into the flat metric of $\mathbbm{R}^4$.

Up to the requisite accuracy, the string frame metric takes the form:
\begin{equation}\label{expandedmetric}
 ds^2=\left(1+\frac{c^2-1}{2}\,\phi ^2+\frac{c-A}{2A}\,\mathbf{y} ^2\right)
 \left[
 \frac{A}{c^2-1}\,dx^2+\frac{1}{A\left(c^2-1\right)^2}\,dc^2 +\frac{1}{A}\,d\phi ^2+\frac{1}{c}\,d\mathbf{y}^2
 \right].
\end{equation}

The $B$-field also contributes to the quadratic part of the action for string fluctuations. This is not immediately evident, because the
coefficients (\ref{aa1}), (\ref{aa2}), (\ref{aa3}) vanish on the classical solution (\ref{classs})-(\ref{class-s5}) and the forms $\sigma _i$ are  transverse to the minimal surface, so the $B$-field seems to vanish on the classical world-sheet. Nevertheless, $\sigma _i$ should be considered of order one, because the $\sigma _i$'s are angular forms on  $S^3$, and $S^3$ shrinks to a point on  the classical solution. As a result, the $B$-field, as a two-form, is actually quadratic in fluctuations.

Taking $\theta \rightarrow 0$ and $\phi =0$ in (\ref{definitionA2}), we find up to the quadratic order in $\theta $:
\begin{equation}
 B=\frac{\sqrt{c^2-1}}{c}\left(\theta d\theta\wedge \sigma _1 +\theta ^2\sigma _2\wedge \sigma _3\right)=\frac{\sqrt{c^2-1}}{2c}\,d\left(\theta ^2\sigma _1\right),
\end{equation}
where we have used (\ref{Maurer-C}) in the second equality. Thus, up to a gauge transformation,
\begin{equation}
 B=\frac{1}{2c^2\sqrt{c^2-1}}\,\theta ^2\sigma _1\wedge dc.
\end{equation}
The Maurer-Cartan forms on $S^3$ can be written as
\begin{equation}
 \sigma _i=\bar{\eta }^i_{mn}n^mdn^n,
\end{equation}
where $\bar{\eta }^i_{mn}$ is the anti-self-dual 't~Hooft symbol \cite{tHooft:1976snw}. Written in the coordinates (\ref{def-y}), the $B$-field becomes
\begin{equation}\label{expandedB}
 B=\frac{1}{2c^2\sqrt{c^2-1}}\,\bar{\eta }_{mn}^1y^mdy^n\wedge dc.
\end{equation}

Expanding (\ref{cveticbosons}) to the quadratic order in fluctuations we get from (\ref{expandedmetric}) and (\ref{expandedB}):
\begin{eqnarray}\label{quadraticLB}
 L_B^{(2)}&=&\frac{1}{2}\,\,\frac{1}{(\sigma ^2-1)^{\frac{3}{2}}}\,\left(\partial _\tau \mathbf{x}\right)^2+\frac{1}{2}\,\frac{A^2}{\sqrt{\sigma ^2-1}}\,\left(\partial _\sigma \mathbf{x}\right)^2
\nonumber \\
&&
 +\frac{1}{2}\,\,\frac{1}{A^2\sqrt{\sigma ^2-1}}\,\left(\partial _\tau \phi \right)^2
 +\frac{1}{2}\,\sqrt{\sigma ^2-1}\left(\partial _\sigma \phi \right)^2
 +\frac{1}{2}\,\,\frac{1}{\sqrt{\sigma ^2-1}}\,\phi ^2
\nonumber \\
&&
+\frac{1}{2}\,\,\frac{1}{A\sigma \sqrt{\sigma ^2-1}}\,\left(\partial _\tau \mathbf{y}\right)^2
+\frac{1}{2}\,\,\frac{A\sqrt{\sigma ^2-1}}{\sigma }\,\left(\partial _\sigma \mathbf{y}\right)^2
+\frac{1}{2}\,\,\frac{\sigma -A}{A(\sigma ^2-1)^{\frac{3}{2}}}\,\mathbf{y}^2
\nonumber \\
&&
+\frac{i}{2\sigma^2 \sqrt{\sigma ^2-1}}\,\bar{\eta }^1_{mn}y^m\partial _\tau y^n,
\end{eqnarray}
where $\mathbf{x}$ is the three-dimensional vector of transverse fluctuations of the string in the 4d space-time directions. In the derivation we used the identities
\begin{align}\label{identitiesAdot}
A' = 2\,\frac{A\sigma -1}{\sigma ^2-1},
\qquad 
 A'' = \frac{{2A}}{{{\sigma ^2} - 1}}\ .
\end{align}
The contributions of the longitudinal modes ($c$ and $x^1$) are cancelled by ghosts. Cancellation of ghost and longitudinal modes is a fairly general phenomenon. We have checked that the respective fluctuation operators are the same by an explicit calculation. The above Lagrangian describes the eight transverse modes of the string.

The fluctuation operators that enter (\ref{generalrationdets}) are defined as
\begin{equation}\label{bosfluctdef}
 S^{(2)}_B=\sum_{{\rm a}}^{}\int_{}^{}d\tau d\sigma \,\sqrt{h}\,\xi ^{\rm a}\mathcal{K}_{\rm a}\xi ^{\rm a},
\end{equation}
 and can be easily read off from (\ref{quadraticLB}). Here $h$ denotes the determinant of the induced world-sheet metric (\ref{wsmetric}):
\begin{equation}\label{sqrth}
 \sqrt{h}=\frac{1}{(\sigma ^2-1)^{\frac{3}{2}}}.
\end{equation}
It is convenient to normalize the fluctuation fields such that the second time derivative has unit coefficient:
\begin{equation}\label{genericK}
 \mathcal{K}=-\partial_\tau^2+\ldots 
\end{equation}
The fields appearing  in (\ref{quadraticLB}) are normalized differently and some field redefinitions are necessary to bring the action into the desired form, which can be achieved by rescaling the fields with appropriate $\sigma $-dependent factors\footnote{These field redefinitions take a simple form after projection of the fluctuations into the local frame $\delta X^{\mu}=E^{\mu}_{\hat a}\xi^{\hat a}$, where the rescaling ${\xi ^{\hat a}} \to \sqrt {\frac{A}{{{\sigma ^2} - 1}}} {\xi ^{\hat a}}$ and partial integration in the action allows us to write the operators in the desired form \eqref{genericK}. This rescaling in the local frame will be compensated by a similar rescaling for fermions, thus preserving the measure of the path integral.}.

After the requisite field redefinitions, we get the following fluctuation Hamiltonians and multiplicities for the three types of modes, $\mathbf{x}$, $\phi $, and $\mathbf{y}$: 
\begin{align}\label{Hamilton}
 \mathcal{K}_\mathbf{x}&=-\partial_\tau ^2-(\sigma ^2-1)^{\frac{3}{2}}\partial _\sigma \,
 \frac{A^2}{\sqrt{\sigma ^2-1}}\,\partial _\sigma,&&\qquad N_\mathbf{x}=3
\nonumber \\
  \mathcal{K}_\phi&=-\partial_\tau ^2-A(\sigma ^2-1)
  \partial _\sigma \sqrt{\sigma ^2-1}\,\partial _\sigma 
   \,\frac{A}{\sqrt{\sigma ^2-1}}
  +A^2,&& \qquad N_\phi=1
\nonumber \\
\mathcal{K}_\mathbf{y}&=\begin{pmatrix}
 \tilde{\mathcal{K}}_\mathbf{y}  &  -\frac{iA}{\sigma }\,\partial_\tau \\ 
 \frac{iA}{\sigma }\,\partial_\tau   &   \tilde{\mathcal{K}}_\mathbf{y} \\ 
 \end{pmatrix},&&\qquad N_\mathbf{y}=2 ,
\end{align}
where
\begin{equation}
  \tilde{\mathcal{K}}_\mathbf{y}  =-\partial_\tau ^2 -\sqrt{{A\sigma }}(\sigma ^2-1)
  \partial _\sigma\,\frac{A\sqrt{\sigma ^2-1}}{\sigma }\,\partial _\sigma  
  \sqrt{\frac{A\sigma }{\sigma ^2-1}}
  +\frac{\sigma \left(\sigma -A\right)}{\sigma ^2-1}\,.
\end{equation}

In deriving the fluctuation operator for the  $\mathbf{y}$-modes, we have used the explicit form of the 't~Hooft symbol:
\begin{equation}
 \bar{\eta }^1_{mn}=\begin{pmatrix}
   0& -1 & 0 & 0 \\ 
  1 & 0 & 0 & 0 \\ 
    0 & 0 & 0 & 1 \\ 
  0 & 0 & -1 & 0 \\ 
 \end{pmatrix}.
\end{equation}
The $\mathbf{y}$-fluctuations decomposed into two identical $2\times 2$ systems upon relabelling of indices. Those can be further disentangled by a similarity transformation:
\begin{align}
U = \frac{1}{{\sqrt 2 }}\left( {\begin{array}{*{20}{c}}
1&i\\
i&1
\end{array}} \right)\ , && {U^\dag }{\mathcal{K}_\mathbf{y}}U = \left( {\begin{array}{*{20}{c}}
{\mathcal{K}_\mathbf{y}^{+}}&0\\
0&{\mathcal{K}_\mathbf{y}^{-}}
\end{array}} \right)\ ,
\end{align}
where
\begin{equation}
 \mathcal{K}_\mathbf{y}^{\pm}=\tilde{\mathcal{K}}_\mathbf{y}\pm \frac{A}{\sigma }\,\partial _\tau .
\end{equation}

Collecting  different pieces together and
using the identities \eqref{identitiesAdot}, we get for the fluctuation operators of the bosonic modes:
\begin{eqnarray}\label{HIs}
 \mathcal{K}_\mathbf{x}&=&-\partial_\tau ^2 -A^2(\sigma ^2-1)\partial _\sigma ^2
 +A\left(4-3A\sigma \right)\partial _\sigma , 
\\
 \mathcal{K}_\phi&=&\mathcal{K}_\mathbf{x}-\frac{2A\sigma }{\sigma ^2-1} ,
 \\
 \label{HIII.}
\mathcal{K}^\pm_\mathbf{y}&=&\mathcal{K}_\mathbf{x}+1-\frac{A\left(\sigma ^2+1\right)\left[4\sigma +3A(\sigma ^2-1)\right]}{4\sigma ^2(\sigma ^2-1)}\pm \frac{A}{\sigma }\,\partial _\tau\,.
\end{eqnarray}
These operators look complicated but are actually related to one another.
 
  The simplest relation is the time reversal symmetry $\tau \rightarrow -\tau $ that maps $\mathcal{K}_\mathbf{y}^+$ to $\mathcal{K}_\mathbf{y}^-$. Since the determinants are time-reversal invariant,
\begin{equation}
 \det\mathcal{K}_\mathbf{y}^-=\det\mathcal{K}_\mathbf{y}^+.
\end{equation}
Another, slightly more intricate relationship connects $\mathcal{K}_\mathbf{x}$ and $\mathcal{K}_\phi$. These operators can be written in a factorized form by introducing the first-order operators
\begin{equation}\label{LbarL}
 {L}=A\sqrt{\sigma ^2-1}\,\partial _\sigma,
 \qquad 
 L^\dagger =-A\sqrt{\sigma ^2-1}\,\partial _\sigma +\frac{2}{\sqrt{\sigma ^2-1}}\,,
\end{equation}
which are Hermitian conjugate with respect to the scalar product
\begin{equation}\label{scprod}
 \left\langle \psi _1\right.\!\!\left|\psi _2 \right\rangle
 =\int_{-\infty }^{+\infty }d\tau \,\int_{1}^{\infty }\frac{d\sigma }{(\sigma ^2-1)^{\frac{3}{2}}}\,\,
 \psi _1^*(\sigma )\psi _2(\sigma )\,.
\end{equation}
It is easy to check that
\begin{equation}\label{factorized}
 \mathcal{K}_\mathbf{x}=-\partial_\tau ^2 +L^\dagger {L},\qquad \mathcal{K}_\phi=-\partial_\tau ^2 +{L}L^\dagger .
\end{equation}
The operators $\mathcal{K}_\mathbf{x}$ and $\mathcal{K}_\phi$, as a consequence, are intertwined by $L$ and ${L}^\dagger $:
\begin{equation}
 \mathcal{K}_\mathbf{x}L^\dagger =L^\dagger \mathcal{K}_\phi,\qquad {L}\mathcal{K}_\mathbf{x}=\mathcal{K}_\phi{L},
\end{equation}
and their eigenfunctions are related: $\psi _\phi\propto L\psi _\mathbf{x}$. The two operators therefore have the same spectra and equal determinants\footnote{For the intertwined operators $\mathcal{K}_\mathbf{x}$ and $ \mathcal{K}_\phi$ to have the same spectra it is also necessary that the map between $\psi_\mathbf{x}$ and $\psi_\phi$ is compatible with the choice of boundary conditions. The latter are discussed in section \ref{spectralproblem}, and by looking at the $\sigma\rightarrow1$ behaviour of the eigenfunctions, we confirmed that this is indeed the case.}:
\begin{equation}\label{K2=K1}
 \det \mathcal{K}_\phi=\det\mathcal{K}_\mathbf{x}.
\end{equation}
The  operators $\mathcal{K}_{\mathbf{x}, \phi}$ are manifestly Hermitian, while $\mathcal{K}_\mathbf{y}^\pm{}^\dagger =\mathcal{K}_\mathbf{y}^\mp$.

With the help of these relationships the bosonic contribution to the partition function can be written as
\begin{equation}
 \det \mathcal{K}_B=\det\nolimits^3\mathcal{K}_\mathbf{x}\det\mathcal{K}_\phi
 \det\nolimits^2\mathcal{K}_\mathbf{y}^+\det\nolimits^2\mathcal{K}_\mathbf{y}^-
 =
 \det\nolimits^4\mathcal{K}_\mathbf{x}\det\nolimits^4\mathcal{K}_\mathbf{y}^+.
\end{equation}

\section{Fermionic fluctuations}\label{fermionsPW}

The fermionic part of the Green-Schwarz action in an arbitrary supergravity background is known explicitly up to second order in fermions \cite{Cvetic:1999zs}. This is enough for our purposes of computing the one-loop contribution to the partition function. After Wick rotation to the Euclidean-signature world-sheet metric, the fermion part of the Lagrangian reads  \cite{Cvetic:1999zs}: 
\begin{eqnarray}\label{cveticfermions}
L_{F}^{(2)} &=&  {{\bar \Psi }^I}\left( {\sqrt h {h^{ij}}{\delta ^{IJ}} + i{\epsilon ^{ij}}\tau _3^{^{IJ}}} \right){\slashed{E}_i}
\left( {{\delta ^{JK}}{D_j} + \frac{{\tau _3^{^{JK}}}}{8}{\partial _j}{X^\nu }{H_{\nu \rho \lambda }}{\Gamma ^{\rho \lambda }} }
\right.
\nonumber \\
&&\left.
+ \frac{{{e^\Phi }}}{8}{\mathcal{F}^{JK}}{\slashed{E}_j} \right){\Psi ^K}.
\end{eqnarray}
The fermion field $\Psi^I$ is a 32-component Majorana-Weyl spinor subject to the constraint $\Gamma ^{11}\Psi ^I=\Psi ^I$.
We use the notations ${\slashed{E}_i} = {\partial _i}{X^\mu }{E_\mu}^{\hat \nu }{\Gamma _{\hat \nu }}$ and ${\Gamma ^{{{\hat \mu }_1}{{\hat \mu }_2}...{{\hat \mu }_n}}} = {\Gamma ^{[{{\hat \mu }_1}}}{\Gamma ^{{{\hat \mu }_2}}}...{\Gamma ^{{{\hat \mu }_n}]}}$, while $D_j$ and ${\mathcal{F}^{JK}}$ are defined by:
\begin{align}
D_j &={\partial _j} + \frac{1}{4}{\partial _j}{X^\mu }{\omega _\mu }^{\hat \alpha \hat \beta }{\Gamma _{\hat \alpha \hat \beta }}\ ,\\
{\mathcal{F}^{JK}} &= \sum\limits_{n = 0}^2 {\frac{1}{{\left( {2n + 1} \right)!}}\tilde{F}_{(2n + 1)}^{{{\hat \mu }_1}{{\hat \mu }_2}...{{\hat \mu }_{2n + 1}}}} {\Gamma _{{{\hat \mu }_1}{{\hat \mu }_2}...{{\hat \mu }_{2n + 1}}}}\ \sigma _{(2n + 1)}^{JK}\ .
\end{align}
Here $\tilde{F}_{(i)}$ are the R-R field strengths, ${\omega _\mu }^{\hat \alpha \hat \beta }$ denotes the spin-connection and $\sigma_{(n)}$ are $2\times2$ matrices defined by:
\begin{align*}
{\sigma _{(1)}} =  - i{\tau _2}\ , &&{\sigma _{(3)}} = \tau_{1}\ , &&{\sigma _{(5)}} =  - \frac{i}{2}{\tau _2}\ .
\end{align*}

The fermionic fluctuation operator is obtained by evaluating the terms of equation \eqref{cveticfermions} that are in between $\bar{\Psi}$ and $\Psi$ on the classical solution (\ref{classs}), (\ref{class-s5}). To do this, we use the field content of the Pilch-Warner background, introduced in sec.~\ref{intropwbackground}, and the following orthonormal frame $E^{\hat{\mu}}$:
\begin{align}
E^{\hat{0}}&\propto dx^{0}, & E^{\hat{1}}&\propto dx^{1}, & E^{\hat{2}}&\propto dx^{2}, & E^{\hat{3}}&\propto dx^{3}, & E^{\hat{4}}&\propto dc, \nonumber\\
E^{\hat{5}}&\propto d\theta, & E^{\hat{6}}&\propto \sigma_{1}, & E^{\hat{7}}&\propto \sigma_{2}, & E^{\hat{8}}&\propto \sigma_{3}, & E^{\hat{9}}&\propto d\phi .
\end{align}
A long but straightforward calculation gives the following expression for the quadratic Lagrangian\footnote{The fermionic operator presented here was calculated using the coordinate $\theta$ of references \cite{Pilch:2000ue,Buchel:2000cn,Pilch:2003jg} for which $\theta_{\rm cl}=\pi/2$, differing from the coordinate used throughout this paper by a shift $\theta\rightarrow\pi/2-\theta$. In principle, both choices have the same physical content as the end result is coordinate independent.}:
\begin{align}\label{32by32fermionicoperator}
{L_{F}^{(2)}} &=  2\sqrt{h}\ \bar{\Psi} \left[ \sqrt {{c_{\left( 1 \right)}}} {\Gamma ^{\hat 1}}{\partial _\tau } + \sqrt {{c_{\left( 2 \right)}}} {\Gamma ^{\hat 4}}{\partial _\sigma } + {c_{\left( \omega  \right)}}{\Gamma ^{\hat 4}} - ic_{\left( 5 \right)}^{\rm{RR}}{\Gamma ^{\hat 0\hat 2\hat 3}}\right.\nonumber\\
&\left. + ic_{\left( 1 \right)}^{\rm{RR}}{\Gamma ^{\hat 1\hat 4\hat 9}} - ic_{\left( 3 \right)}^{\rm{NSNS}}\left( {{\Gamma ^{\hat 1\hat 5\hat 6}} - {\Gamma ^{\hat 1\hat 7\hat 8}}} \right) + c_{\left( 3 \right)}^{\rm{RR}}\left( {{\Gamma ^{\hat 5\hat 6\hat 9}} - {\Gamma ^{\hat 7\hat 8\hat 9}}} \right) \right]\Psi,
\end{align}
where the coefficients are
\begin{align}
{c_{\left( 1 \right)}} &= \frac{{{\sigma ^2} - 1}}{A}\,,&
{c_{\left( 2 \right)}} &= A{\left( {{\sigma ^2} - 1} \right)^2},
\nonumber\\
{c_{\left( \omega  \right)}} &=  - \frac{1}{{2\sqrt A }}\,,&
c_{\left( 1 \right)}^{\rm{RR}} &=  - \frac{1}{{4\sigma }}\sqrt A \left( {{\sigma ^2} - 1} \right),
\nonumber\\
c_{\left( 3 \right)}^{\rm{RR}} &=  - \frac{{\left( {2\sigma  + A} \right)\sqrt {{\sigma ^2} - 1} }}{{4\sigma \sqrt A }}\,,&
c_{\left( 3 \right)}^{\rm{NSNS}} &= \frac{{\sqrt {A\left( {{\sigma ^2} - 1} \right)} }}{{4\sigma }},
\nonumber\\
c_{\left( 5 \right)}^{\rm{RR}} &= \frac{{4\sigma  - \left( {{\sigma ^2} - 1} \right)A}}{{4\sigma \sqrt A }}\,.\nonumber
\end{align}
We used the identities \eqref{identitiesAdot} and the positive chirality condition ${\Gamma ^{\hat 0\hat 1\hat 2\hat 3\hat 4}}\Psi = {\Gamma ^{\hat 5\hat 6\hat 7\hat 8\hat 9}}\Psi$ in the course of the derivation.
The $\kappa$-symmetry gauge-fixing condition is the same as  in \cite{Drukker:2000ep,Kruczenski:2008zk}: $\Psi^{1}=\Psi^{2}=\Psi$.  Our conventions for the ten-dimensional Dirac algebra are summarized in appendix~\ref{conventionsandnotations}.

The first two terms in \eqref{32by32fermionicoperator} come from the kinetic terms in the fermionic Lagrangian, the third term originates from the spin-connection,  the fourth term corresponds to the contribution of the R-R 5-form $\tilde{F}_{(5)}$. The terms in the second line correspond to the contributions of the R-R 1-form $\tilde{F}_{(1)}$, the NS-NS field strength $H$, and the R-R field strength $\tilde{F}_{(3)}$.

The $\mathfrak{so}(4,2)$-plus-$\mathfrak{so}(6)$ decomposition of the Dirac matrices described in the appendix~\ref{conventionsandnotations}, yields the following form of the fermionic Lagrangian:
\begin{align}\label{16by16fermionicoperator}
{L_{F}^{(2)}} =&  2\sqrt h \;\bar \chi \left[ \sqrt {{c_{\left( 1 \right)}}} {\gamma ^{\hat 1}}{\partial _\tau } + \sqrt {{c_{\left( 2 \right)}}} {\gamma ^{\hat 4}}{\partial _\sigma } + {c_{\left( \omega  \right)}}{\gamma ^{\hat 4}} - c_{\left( 5 \right)}^{\rm{RR}}{\gamma ^{\hat 1\hat 4}}\right.\nonumber\\
&\left. - c_{\left( 1 \right)}^{\rm{RR}}{\gamma ^{\hat 1\hat 4\hat 9}} - ic_{\left( 3 \right)}^{\rm{NSNS}}\left( {{\gamma ^{\hat 1\hat 5\hat 6}} - {\gamma ^{\hat 1\hat 7\hat 8}}} \right) + ic_{\left( 3 \right)}^{\rm{RR}}\left( {{\gamma ^{\hat 5\hat 6\hat 9}} - {\gamma ^{\hat 7\hat 8\hat 9}}} \right) \right]\chi,
\end{align}
where $\chi$ is a 16-component spinor and the various terms are written in the same order as in \eqref{32by32fermionicoperator}. We explicitly checked in appendix~\ref{AppendexComments} that taking the near-boundary limit:  $\sigma\rightarrow 1+{z^2}/{2}$, and keeping only the leading terms in $z$, we recover the quadratic action for the string in $AdS_5\times S^5$ from \cite{Drukker:2000ep,Kruczenski:2008zk}.

The fermionic Lagrangian can be simplified by judicious choice of representation of the Dirac matrices. We take the following representation for the $4\times4$ Dirac matrices $\gamma^{\hat{a}}$ and $\gamma^{\hat{a}'}$ described in appendix \ref{conventionsandnotations}
\begin{align}\label{choiceofgammas}
{\gamma ^{\hat 0}} &= i{\tau _2} \otimes {\tau _1} , &
{\gamma ^{\hat 1}} &=  - {\tau _3} \otimes \mathbbm{1} , &
{\gamma ^{\hat 2}} &= {\tau _2} \otimes {\tau _2} , &
{\gamma ^{\hat 3}} &= {\tau _2} \otimes {\tau _3} , &
{\gamma ^{\hat 4}} &= {\tau _1} \otimes \mathbbm{1}\ ,\nonumber\\
{\gamma ^{\hat 5'}} &= {\gamma ^{\hat 4}} ,&
{\gamma ^{\hat 6'}} &= {\gamma ^{\hat 3}},&
{\gamma ^{\hat 7'}} &= {\gamma ^{\hat 2}} ,&
{\gamma ^{\hat 8'}} &= i{\gamma ^{\hat 0}},&
{\gamma ^{\hat 9'}} &= {\gamma ^{\hat 1}}.
\end{align}
This choice  is by no means unique. However, it allows us to decompose the fermionic operator in terms of $2\times2$ operators, instead of more complicated $4\times4$ operators that one would be left with in a generic representation of the $\mathfrak{so}(6)/\mathfrak{so}(4,2)$ Clifford algebra. 

As in the case of bosons, we rescale the fluctuation fields in order to normalize the coefficient in front of $\partial_{\tau}$ to one. The requisite rescaling is
\begin{equation}
{\chi } \to \frac{1}{{c_{\left( 1 \right)}^{1/4}}}\ {\psi }.
\end{equation}
 After the rescaling, the fermionic Lagrangian can be brought to the following form with the help of eqs.~\eqref{identitiesAdot}:
\begin{eqnarray}\label{fermionicactionPWfinal}
{L_{F}^{(2)}} &=& 2\sqrt h \left[ \sum\limits_{j = 1}^4 \left( {\begin{array}{*{20}{c}}
{{{\bar \psi }_{2j - 1}}}&{{{\bar \psi }_{2j}}}
\end{array}} \right)\tau_3{\mathcal{D}_{0}}\left( {\begin{array}{*{20}{c}}
{{\psi _{2j - 1}}}\\
{{\psi _{2j}}}
\end{array}} \right) 
\right.\nonumber\\
&&\left. 
+ \sum\limits_{j = 5}^6 \left( {\begin{array}{*{20}{c}}
{{{\bar \psi }_{2j - 1}}}&{{{\bar \psi }_{2j}}}
\end{array}} \right)\tau_3{\mathcal{D}_+}\left( {\begin{array}{*{20}{c}}
{{\psi _{2j - 1}}}\\
{{\psi _{2j}}}
\end{array}} \right) 
\right.\nonumber\\
&&\left. 
+ \sum\limits_{j = 7}^8 {\left( {\begin{array}{*{20}{c}}
{{{\bar \psi }_{2j - 1}}}&{{{\bar \psi }_{2j}}}
\end{array}} \right)\tau_3{\mathcal{D}_{-}}\left( {\begin{array}{*{20}{c}}
{{\psi _{2j - 1}}}\\
{{\psi _{2j}}}
\end{array}} \right)}    \right],
\end{eqnarray}
where:
\begin{eqnarray}
{\mathcal{D}_{0}} &=& \left( {\begin{array}{*{20}{c}}
{{\partial _\tau }}&{A\sqrt {\sigma ^2-1}\, {\partial _\sigma }  - \frac{2}{{\sqrt {{\sigma ^2} - 1} }}}\\
{-A\sqrt {\sigma ^2-1} \,{\partial _\sigma } }&{  {\partial _\tau }}
\end{array}} \right) , 
\label{D4} \\
 \label{D+-}
{\mathcal{D}_ \pm } &= &\left( {\begin{array}{*{20}{c}}
{{\partial _\tau } \pm 1 \pm \frac{A}{\sigma }}&{A\sqrt {\sigma ^2-1}\, {\partial _\sigma } + \frac{{\left( {{\sigma ^2} - 1} \right)A - 4\sigma  }}{{2\sigma \sqrt {{\sigma ^2} - 1} }}}\\
{-A\sqrt {\sigma ^2-1}\, {\partial _\sigma } + \frac{{A\sqrt {{\sigma ^2} - 1} }}{{2\sigma }}}&{ {\partial _\tau } \mp 1}
\end{array}} \right),
\end{eqnarray}

The operators $\mathcal{D}_\pm$ are related by time reversal:
\begin{equation}
 \left.\vphantom{\frac{1}{2}}
 \mathcal{D}_\pm\right|_{\tau \rightarrow -\tau }
 =-\tau_3\mathcal{D}_\mp\tau_3,
\end{equation}
so $\det\mathcal{D}_+=\det\mathcal{D}_-$, and we get for
the fermionic partition function:
\begin{equation}
 \det \mathcal{K}_F=\det\nolimits^4\mathcal{D}_{0}\det\nolimits^2\mathcal{D}_+
 \det\nolimits^2\mathcal{D}_-=\det\nolimits^4\mathcal{D}_{0}\det\nolimits^4\mathcal{D}_-.
\end{equation}

\section{The semiclassical partition function}\label{section6}

When comparing fermionic and bosonic contributions to the partition function, we first notice that the Dirac operator $\mathcal{D}_{0}$ is built from the same intertwiners (\ref{LbarL}) that appear in the analysis of the bosonic modes:
\begin{equation}
 \mathcal{D}_{0}=\begin{pmatrix}
  \partial _\tau  & -{L}^\dagger  \\ 
 -L  & \partial _\tau  \\ 
 \end{pmatrix}.
\end{equation}
Squaring the Dirac operator, we find:
\begin{equation}\label{D4-K1-K2}
 \left(\tau_3\mathcal{D}_{0}\right)^2=-\begin{pmatrix}
  \mathcal{K}_\mathbf{x} & 0  \\ 
  0 & \mathcal{K}_\phi \\ 
 \end{pmatrix},
\end{equation}
which follows from the factorized representation (\ref{factorized}) of the bosonic fluctuation operators
$\mathcal{K}_\mathbf{x}$ and $\mathcal{K}_\phi$. Since $\mathcal{K}_\mathbf{x}$ and $\mathcal{K}_\phi$ are isospectral, $\det^4\mathcal{D}_{0}=\det^2\mathcal{K}_\mathbf{x}\det^2\mathcal{K}_\phi=\det^4\mathcal{K}_\mathbf{x}$, and the contribution of these operators cancels between bosons and fermions\footnote{We assume that the spectrum of $\mathcal{K}_\mathbf{x}$ and $\mathcal{K}_{\phi}$ is the same when appearing in bosons and fermions. This is a consequence of choosing the same boundary conditions in both cases. The prescription for the latter will be explained in section \ref{spectralproblem}.}:
\begin{equation}\label{rationdets-not-final}
 W(C)=\,{\rm e}\,^{-S_{\rm cl}}\,\frac{\det^2\mathcal{D}_-}{\det^{2}\mathcal{K}_\mathbf{y}^+}\,.
\end{equation}

These cancellations are very suggestive, and call for introducing another pair of intertwiners:
\begin{equation}
 \mathcal{L}={A\sqrt {\sigma ^2-1}\, {\partial _\sigma } - \frac{{A\sqrt {{\sigma ^2} - 1} }}{{2\sigma }}}\,,
 \qquad 
 \mathcal{L}^\dagger =-{A\sqrt {\sigma ^2-1}\, {\partial _\sigma } 
  - \frac{{A\sqrt {{\sigma ^2} - 1} }}{{2\sigma }}}
  +\frac{2}{\sqrt{\sigma ^2-1}}\,,
\end{equation}
which are also conjugate with respect to the scalar product (\ref{scprod}). The Dirac operator (\ref{D+-}) then takes the form:
\begin{equation}\label{Dirac-massive}
 \mathcal{D}_\pm=\begin{pmatrix}
 \partial _\tau \pm 1\pm\frac{A}{\sigma }& -\mathcal{L}^\dagger  \\ 
  -\mathcal{L} & \partial _\tau \mp 1 \\ 
 \end{pmatrix}.
\end{equation}

The operators $\mathcal{K}_\mathbf{y}^\pm$ can also be neatly expressed through $\mathcal{L}$, $\mathcal{L}^\dagger $:
\begin{equation}\label{intertKIII}
 \mathcal{K}_\mathbf{y}^\pm=-\partial _\tau ^2+\mathcal{L}\mathcal{L}^\dagger 
 +\frac{A}{\sigma }+1\pm\frac{A}{\sigma }\,\partial _\tau .
\end{equation}
Using the formula for the determinant of a block matrix:
\begin{equation}
 \det\begin{pmatrix}
 A  & B \\ 
  C & D \\ 
 \end{pmatrix}
 =\det\left(AD-BD^{-1}CD\right)
 \stackrel{{\rm if~}[C,D]=0}{=}\det\left(AD-BC\right),
\end{equation}
the determinant of the Dirac operator (\ref{Dirac-massive}) can be brought to the second-order scalar form:
\begin{equation}\label{secordferm}
 \det\mathcal{D}_\pm=\det\left(-\partial _\tau ^2+
 \mathcal{L}^\dagger \mathcal{L}+\frac{A}{\sigma } + 1 \mp\frac{A}{\sigma }\,\partial _\tau \right),
\end{equation}
which is very similar to (\ref{intertKIII}),  but not entirely identical. The two operators differ by the order in which intertwiners are multiplied. They are not isospectral to one another because of the extra $\sigma $-dependent terms in the potential proportional to $A/\sigma $. 

The second-order form of a Dirac determinant is typically more convenient for practical calculations. Here we found, on the contrary,  the first-order matrix form much easier to deal with. Its practical convenience stems from the simple dependence on the time derivative. The second-order form (\ref{secordferm}) contains the time derivative multiplied by a $\sigma $-dependent term, which substantially complicates the analysis. We thus keep the fermion operator in its original Dirac form.

 Moreover, it is useful to rewrite the bosonic determinant in the first-order form as well:
\begin{equation}
\det\mathcal{K}_\mathbf{y}^\mp=\det\begin{pmatrix}
 \partial _\tau \pm 1\pm\frac{A}{\sigma }& -\mathcal{L}  \\ 
  -\mathcal{L}^\dagger  & \partial _\tau \mp 1 \\ 
 \end{pmatrix}.
\end{equation}
By introducing two Dirac-type Hamiltonians:
\begin{equation}\label{basicDirac1}
 \mathcal{H}_B=\begin{pmatrix}
  1+\frac{A}{\sigma }  & \mathcal{L} \\ 
  \mathcal{L}^\dagger  & -1 \\ 
 \end{pmatrix},\qquad 
  \mathcal{H}_F=\begin{pmatrix}
  -1                   & \mathcal{L}  \\ 
  \mathcal{L}^\dagger  & 1+\frac{A}{\sigma }  \\ 
 \end{pmatrix},
\end{equation}
we can bring (\ref{rationdets-not-final}) to the form:
\begin{equation}
 \label{rationdets-final}
 W(C)=\,{\rm e}\,^{-S_{\rm cl}}\,\frac{\det^2\left(\partial _\tau -\mathcal{H}_F\right)}{\det^2\left(\partial _\tau -\mathcal{H}_B\right)}\,.
\end{equation}
We have  performed an innocuous similarity transformation with the first Pauli matrix to the fermion operator \eqref{Dirac-massive}.
This expression will be our starting point for the evaluation of the one-loop correction to the Wilson loop expectation value.

\subsection{Spectral problem}\label{spectralproblem}

The Fourier transform eliminates the $\tau $-dependence in the determinants: 
\begin{equation}\label{afterFourier}
 \ln\det\left(\partial _\tau -\mathcal{H}\right)={\rm L}\int_{-\infty }^{+\infty }\frac{d\omega }{2\pi }\,\,\mathop{\mathrm{tr}}\ln\left(i\omega-\mathcal{H} \right),
\end{equation}
leaving us with a one-dimensional problem of finding the spectra of the Dirac operators (\ref{basicDirac1}):
\begin{equation}\label{sch1}
 \mathcal{H}\psi =E\psi .
\end{equation}

The Dirac operators are Hermitian with respect to the scalar product (\ref{scprod}) and consequently have real eigenvalues. The measure factor
in the scalar product originates from the induced metric on the world-sheet, as it appears in (\ref{bosfluctdef}), (\ref{sqrth}). Alternatively, one can absorb the measure into the wavefunction:
\begin{equation}
 \psi =(\sigma ^2-1)^{\frac{3}{4}}\chi .
\end{equation}
The resulting eigenvalue problem,
\begin{equation}\label{sch2}
 \hat{\mathcal{H}}\chi =E\chi 
\end{equation}
is Hermitian with respect to the conventional scalar product without any measure factors. The Dirac operators $\hat{\mathcal{H}}_{B,F}$ have the same form as (\ref{basicDirac1}) but with transformed $\mathcal{L}$, $\mathcal{L}^\dagger $, i.e.
\begin{equation}\label{basicDirac}
 \hat{\mathcal{H}}_B=\begin{pmatrix}
  1+\frac{A}{\sigma }  & \hat{\mathcal{L}} \\ 
  \hat{\mathcal{L}}^\dagger  & -1 \\ 
 \end{pmatrix},\qquad 
  \hat{\mathcal{H}}_F=\begin{pmatrix}
  -1                   & \hat{\mathcal{L}} \\ 
  \hat{\mathcal{L}}^\dagger  & 1+\frac{A}{\sigma }  \\ 
 \end{pmatrix},
\end{equation}
with
\begin{eqnarray}\label{EuclideanLs}
 \hat{\mathcal{L}}&=&A\sqrt{\sigma ^2-1}\,\partial _\sigma +\frac{A\left(2\sigma ^2+1\right)}{2\sigma \sqrt{\sigma ^2-1}} ,
\nonumber \\
\hat{\mathcal{L}}^\dagger &=&-A\sqrt{\sigma ^2-1}\,\partial _\sigma 
-\frac{A\left(4\sigma ^2-1\right)}{2\sigma \sqrt{\sigma ^2-1}}
+\frac{2}{\sqrt{\sigma ^2-1}}\,.
\end{eqnarray}

\subsubsection{Boundary conditions}
The Dirac equation  (\ref{sch2}) must be supplemented with boundary conditions at $\sigma =1$ and $\sigma \rightarrow \infty $. Near the boundary,
\begin{equation}
 A= 1+\mathcal{O}\left((\sigma -1)\ln(\sigma -1)\right)\qquad \left(\sigma \rightarrow 1\right),
\end{equation}
and the Dirac operators (minus the eigenvalue) asymptote to
\begin{equation}\label{asymDirac}
 \hat{\mathcal{H}}_{B,F} - E  =\begin{pmatrix}
  0  & \sqrt{2\left(\sigma -1\right)}\,\partial _\sigma +\frac{3}{\sqrt{8\left(\sigma -1\right)}} \\ 
  -\sqrt{2\left(\sigma -1\right)}\,\partial _\sigma  +\frac{1}{\sqrt{8\left(\sigma -1\right)}}  & 0 \\ 
 \end{pmatrix} + \mathcal{O}(1)
\end{equation}

By requiring the right-hand side of \eqref{asymDirac} to vanish when applied to the wavefunction ansatz (proportional to a constant vector)
\begin{equation}
 \chi_{B, F} \propto  (\sigma-1)^\nu,
\end{equation}
two solutions are found:
\begin{eqnarray}
 \chi_{B, F} &\simeq &C_{B, F}^-\left(\sigma -1\right)^{-\frac{3}{4}}\begin{pmatrix}
  0 \\ 
  1 \\ 
 \end{pmatrix}
 +C_{B, F}^+\left(\sigma -1\right)^{\frac{1}{4}}\begin{pmatrix}
  1 \\ 
  0 \\ 
 \end{pmatrix} 
\qquad \left(\sigma \rightarrow 1\right).
\end{eqnarray}
The Dirichlet boundary conditions for the string fluctuations require  the growing, non-normalizable solution to be absent:
\begin{equation}\label{b.c.1}
 C^-=0.
\end{equation}
This condition fixes the solution uniquely, up to an overall normalization, which can be further fixed by setting $C^+=1$.
In conclusion, the leading close-to-boundary behaviour for the (normalizable) eigenfunction is:
\begin{eqnarray}\label{sigmato1solutions}
 \chi_{B, F} &\simeq
\left(\sigma -1\right)^{\frac{1}{4}}\begin{pmatrix}
  1 \\ 
  0 \\ 
 \end{pmatrix} 
\qquad \left(\sigma \rightarrow 1\right).
\end{eqnarray}

At large $\sigma $,
\begin{equation}
 A=\frac{2}{3\sigma }+\mathcal{O}\left(\frac{1}{\sigma ^3}\right)
 \qquad 
 \left(\sigma \rightarrow \infty \right).
\end{equation}
The potential terms in the intertwiners vanish at infinity,
\begin{equation}
 \hat{\mathcal{L}}\simeq \frac{2}{3}\,\partial _\sigma \simeq -\hat{\mathcal{L}}^\dagger \qquad \left(\sigma \rightarrow \infty \right),
\end{equation}
and (\ref{basicDirac}) become free, massive Dirac operators. 

The eigenvalue problem (\ref{sch2}) thus describes a one-dimensional relativistic fermion bouncing off an infinite wall at $\sigma =1$. The spectrum of this problem is continuous and non-degenerate. There are two branches corresponding to particles and holes. Each particle or hole state can be labelled by the asymptotic value of the momentum $p\in [0,\infty )$, in terms of which the eigenvalue is given by 
\begin{equation}\label{eigenE}
 E=\pm\sqrt{\frac{4}{9}\,p^2+1}\,.
\end{equation}
The positive-energy eigenstates correspond to particles and the negative-energy ones to holes.

The asymptotic wavefunctions are plane waves:
\begin{eqnarray}\label{boundaryCs}
 \chi _B&\simeq& C_B^\infty \begin{pmatrix}
  \sin \left(p\sigma +\delta _B^\pm\right) \\ 
  -\dfrac{2 p }{3 (\pm |E|+1)}\cos\left(p\sigma +\delta _B^\pm\right) \\
 \end{pmatrix},
\nonumber \\
 \chi _F&\simeq&C_F^\infty \begin{pmatrix}
    \sin \left(p\sigma +\delta _F^\pm\right) \\ 
   - \dfrac{2 p }{3 (\pm |E|-1)}\cos\left(p\sigma +\delta _F^\pm\right) \\
 \end{pmatrix} 
 \qquad \left(\sigma \rightarrow \infty \right),
\end{eqnarray}
where $\delta_{B,F}^\pm \equiv \delta_{B,F}^\pm (p)$ are the phaseshifts experienced by particles/holes as they reflects from the wall at $\sigma =1$. Since the particle-hole symmetry is broken by the $A/\sigma $ term in the Dirac Hamiltonian, particles and holes have different phaseshifts: 
$\delta^+ (p)\neq\delta ^-(p)$. 

\subsection{Phaseshifts}

The density of states in the continuum and with it the operator determinants are usually expressed through the scattering phaseshifts. This relation is routinely used in soliton quantization \cite{Dashen:1974cj,tHooft:1976snw}. Let us briefly recall the standard argument. To regulate the problem we can impose fiducial boundary conditions at some large $\sigma =R$. For instance,
\begin{equation}\label{artificialBC}
 \left(1+\tau_3\right)\chi (R)=0.
\end{equation}
This makes the spectrum discrete.
Taking into account the asymptotic form of the wavefunction (\ref{boundaryCs}), the boundary condition leads to momentum quantization:
\begin{equation}
 p_nR+\delta (p_n)=\pi n,
\end{equation}
from which we find the density of states:
\begin{equation}
 \rho (p)=\frac{dn}{d p}=\frac{R}{\pi }+\frac{1}{\pi }\,\,\frac{d\delta(p) }{dp}\,.
\end{equation}
The leading-order constant term in the density of states gives an extensive contribution to the partition function, proportional to the internal length of the string, but this will cancel  in the ratio of determinants (\ref{rationdets-final}). We can thus ignore the constant term and concentrate on the $O(1)$ momentum-dependent distortion due to the phaseshift.

Taking into account  (\ref{eigenE}), we rewrite (\ref{afterFourier}) as 
\begin{eqnarray}
 \ln\det\left(\partial _\tau -\mathcal{H}\right)&=&{\rm{L}}\int_{-\infty }^{+\infty }\frac{d\omega }{2\pi }\,
 \int_{0}^{\infty }\frac{dp}{\pi }\,\,
 \left(
 \frac{d\delta^+ (p)}{d p}
 \ln\left(i\omega -\sqrt{\frac{4}{9}\,p^2+1}\right)
 \right.
\nonumber \\
 &&\left.
 +
 \frac{d\delta^- (p)}{dp}
 \ln\left(i\omega +\sqrt{\frac{4}{9}\,p^2+1}\right)
 \right).
\end{eqnarray}
Integration by parts gives
\begin{eqnarray}
 \ln\det\left(\partial _\tau -\mathcal{H}\right)&=&{\rm{L}}\,\frac{4}{9}\int_{-\infty }^{+\infty }\frac{d\omega }{2\pi }\,
 \int_{0}^{\infty }\frac{dp}{\pi }\,\,
 \frac{p}{\sqrt{\frac{4}{9}\,p^2+1}}
 \left(
 \frac{\delta ^+(p)}{i\omega -\sqrt{\frac{4}{9}\,p^2+1}}
 \right.
\nonumber \\
&&\left.
 -
 \frac{\delta ^-(p)}{i\omega +\sqrt{\frac{4}{9}\,p^2+1}}
 \right).
\end{eqnarray}
The integral over $\omega $ is a half-residue at infinity and we finally obtain:
\begin{equation}\label{ldetfinal}
 \ln\det\left(\partial _\tau -\mathcal{H}\right)=-{\rm{L}}\,\frac{4}{9}
 \int_{0}^{\infty }\frac{dp}{2\pi }\,\,
  \frac{p}{\sqrt{\frac{4}{9}\,p^2+1}}
 \left({\delta ^+(p)} +
{\delta ^-(p)} \right).
\end{equation}

The effective string tension, as defined in (\ref{perlaw}), is minus the free energy per unit length. We can write:
\begin{equation}
 T(\lambda )=\frac{\sqrt{\lambda }}{2\pi }-\frac{1}{4}-\frac{\Delta }{4} ,
\end{equation}
 where the second term comes from the dilaton coupling through the Fradkin-Tseytlin term, and the last term is the genuine quantum contribution of string fluctuations. Using (\ref{ldetfinal}) to express the determinants in  (\ref{rationdets-final}) through phaseshifts  we get:
\begin{equation}\label{deltaphases}    \boxed{
 \Delta =\frac{32}{9}\, \int_{0}^{\infty }\frac{dp}{2\pi }\,\,
 \frac{p }{\sqrt{\frac{4}{9}\,p^2+1}}\,\left(
 \delta_F ^+(p)
 +
\delta_F ^-(p)
-\delta_B ^+(p)
 -
\delta_B^-(p)
\right)}\, .
\end{equation}
The large-$N$ localization predicts $\Delta =1$, as seen from eq.~(\ref{T(lambda)}). We are going to compute $\Delta $ on the string side of the duality by numerically evaluating the phaseshifts entering (\ref{deltaphases}).

It is easy to convince oneself, for instance using the WKB approximation for the wavefunctions, that the phaseshifts grow linearly at large momenta. The momentum integral in (\ref{deltaphases}) therefore is potentially divergent. This is not surprising since individual loop integrals in the 2d sigma-model that defines the string path integral are UV divergent. The supergravity equations of motion however should guarantee that the divergences cancel and the complete result is UV finite, at least in the one-loop approximation. Cancellation of divergences is a strong consistency check on our calculations, since the fermionic and bosonic phaseshifts should compensate one another up to the $\mathcal{O}(1/p^2)$ accuracy. In other words, the first three orders of the $1/p$ expansion should cancel. The large-$p$ expansion of the phaseshifts is essentially equivalent to the WKB expansion for the wavefunctions, which we carry out to the requisite order in the appendix~\ref{large-p}, where we show that the divergences cancel out as expected.

Another check on our formalism is to see that in $AdS_5\times S^5$ the quantum string correction to the expectation value of the straight Wilson line vanishes. The PW geometry asymptotes to $AdS_5\times S^5$ near the boundary, and the AdS result can be viewed as a limiting case of our calculation where the near-boundary limit of the fluctuation operators is taken first, prior to computing the phaseshifts (see appendix~\ref{AppendexComments}). The $AdS_5\times S^5$ fluctuation problem is sufficiently simple and all the phaseshifts can be found analytically. We show in the appendix~\ref{TheAds5xS5Case} that the bosonic and fermionic phaseshifts conspire to cancel at the level of integrand, demonstrating that indeed the straight Wilson line is not renormalized in $AdS_5\times S^5$.

\subsection{Numerics}

Although the bosonic and fermionic operators in \eqref{basicDirac} look enticingly similar,
we were so far unable to solve the spectral problem \eqref{sch2} analytically. 
Thus, we resort to numerics in order to evaluate $\Delta$. 

The idea is that, first, we numerically solve the different spectral problems with the conditions \eqref{sigmato1solutions} at $\sigma \rightarrow 1$, and then numerically evolve the wavefunctions far away from the boundary,
which is the phaseshift regime.
Then, we fit the resulting asymptotic eigenfunctions to plane waves and find their phaseshifts. 
This procedure is done for a range of values in $p$, but not for $p=0$ since the solution would not be oscillatory.
Finally, we integrate numerically over $p$ to evaluate \eqref{deltaphases}.
The numeric parameters used are presented in appendix \ref{sec:numericError}.

\begin{figure}[h]\label{plotdeltas}
\includegraphics[scale=0.53]{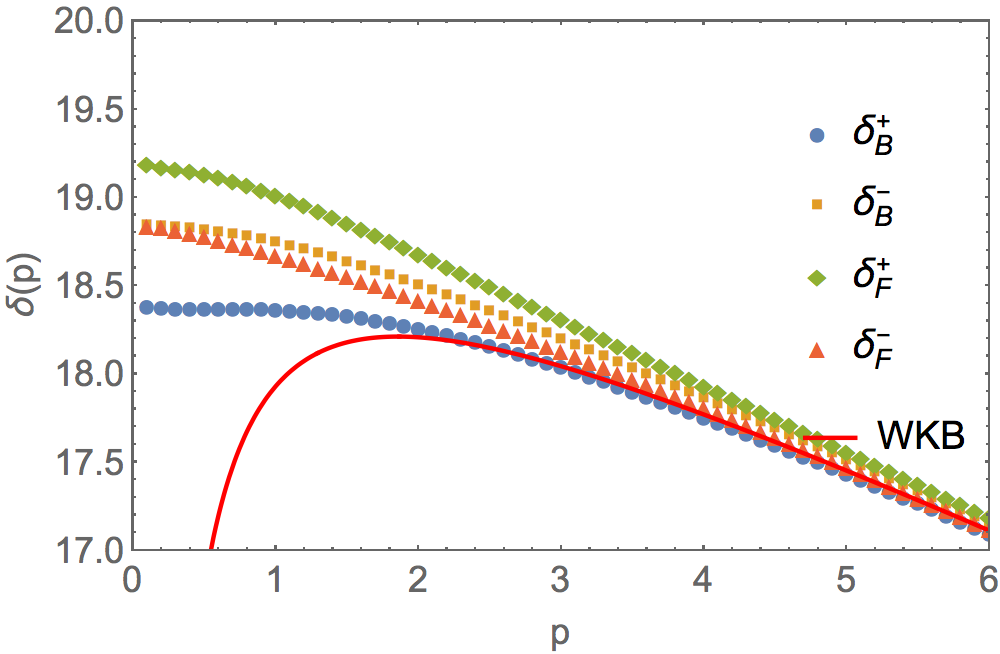}
\includegraphics[scale=0.52]{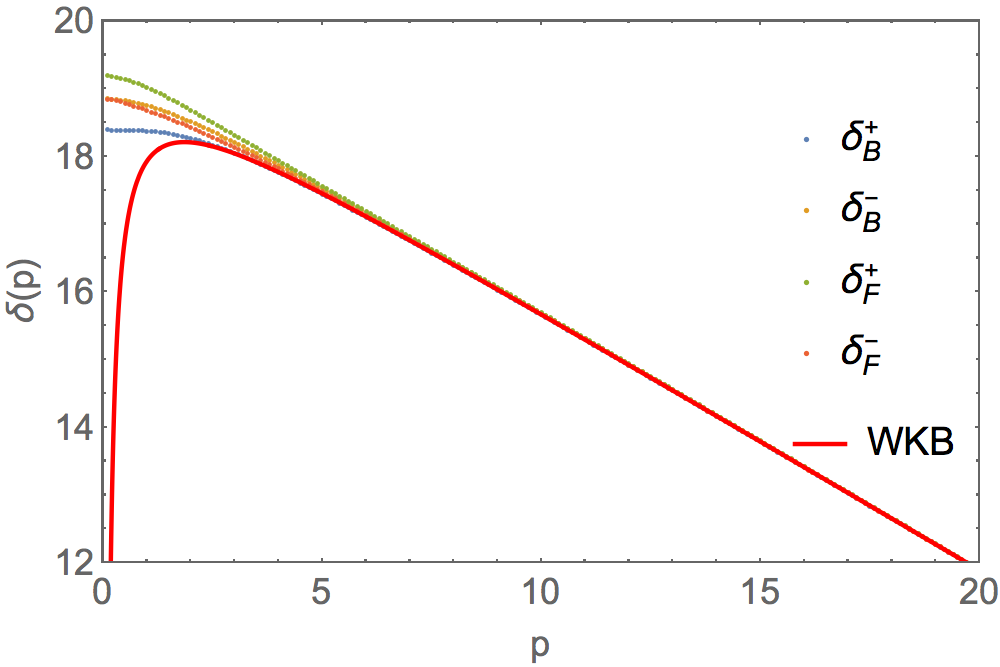}
\centering
\caption{Numerical results for the phaseshifts and the WKB approximation \eqref{WKBphaseshift} as functions of $p$. }
\label{fig:Phaseshifts}
\end{figure}

Our algorithm measures phaseshifts up to a constant, which we recall does not contribute to our ratio of determinants.  
The four phaseshifts (constant-shifted to match the same asymptotics) associated to the operators $\hat{\mathcal{H}}_{B,F}$, 
with positive and negative energy, are plotted in figure \ref{fig:Phaseshifts}. 
In the latter, the WKB approximation common for all the phaseshifts \eqref{WKBphaseshift} is also shown, displaying  nice agreement for large $p$.

\begin{figure}[h]
\includegraphics[scale=0.8]{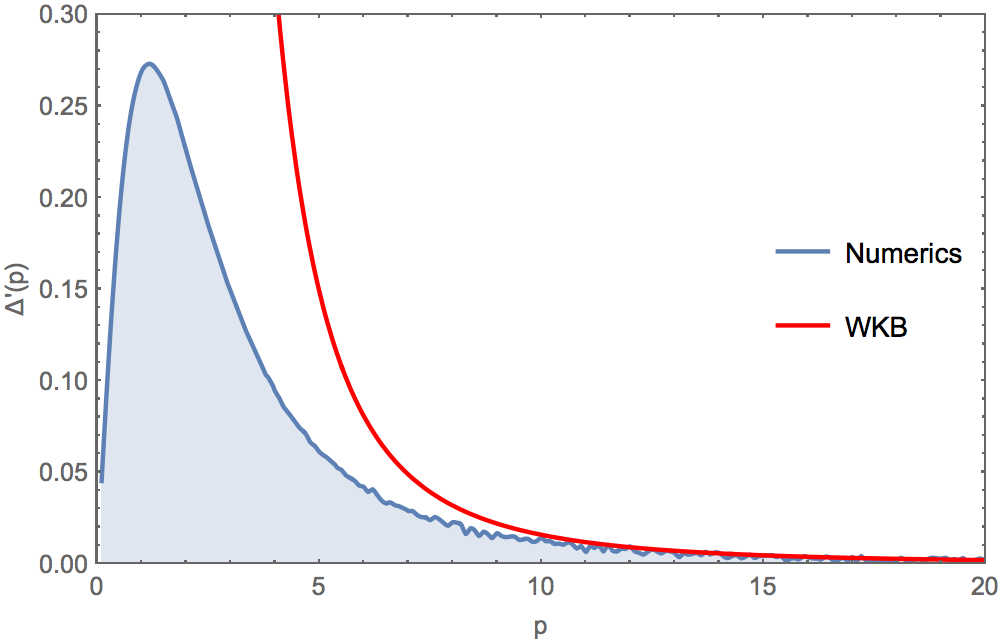}
\centering
\caption{The integrand $\Delta'$ as a function of $p$ and the corresponding WKB result from \eqref{differencePhaseshiftsWKB}. The area under the curve, $\Delta$, is given in equation \eqref{FinalDelta}. }
\label{fig:IntegrandFig}
\end{figure}

We are interested though in the difference of phaseshifts, or more precisely in the integrand of \eqref{deltaphases}. 
In figure \ref{fig:IntegrandFig}, we plot the integrand resulting from numerics as a function of $p$, together with the corresponding expression from the WKB approximation \eqref{differencePhaseshiftsWKB}. 
Indeed, as predicted by WKB, cancellation of phaseshifts is observed for large $p$, thus making the area under the curve, and with it $\Delta$, a finite quantity.

Finally, the numerical integration returns a result that matches with the prediction from localization, within the numerical error (see appendix \ref{sec:numericError} for the error estimate):
\begin{equation}\label{FinalDelta}
\boxed{\Delta = 1.01 \pm 0.03} \,.
\end{equation}

\section{Conclusions}

We found complete agreement between the exact prediction from the field theory, extrapolated to strong coupling, and an explicit string-theory calculation of the effective string tension. This result provides another quantitative test of the $\mathcal{N}=2^*$ holography. The quantity that we calculated can be regarded as a holographic counterpart of the L\"uscher correction, and requires fully quantum mechanical treatment of the string world-sheet. Our calculations demonstrate that string theory in the PW background can be consistently quantized, despite the background's complexity and its reduced degree of supersymmetry. It also elucidates the role of the Fradkin-Tseytlin term in the Green-Schwarz formalism. The ensuing dilaton coupling was necessary to bring the result of the string calculation in agreement with the field-theory predictions. 

The field-theory predictions for the effective string tension have been originally obtained by taking the infinite-radius limit of the circular Wilson loop on $S^4$, which can be calculated exactly with the help of localization.  The supergravity background with the $S^4$ boundary is actually known, and is better-behaved in the IR, but only as a solution of the 5d Einstein's equations \cite{Bobev:2013cja}. In order to consistently define the string action on this background it is first necessary to uplift the solution to ten dimensions. 

Finally, it would be interesting to generalize our calculations to other string solutions in the Pilch-Warner background \cite{Dimov:2003bh,Buchel:2013id,Young:2014jma},
to the string dual of the pure $\mathcal{N}=2$ theory \cite{Bigazzi:2001aj,Bigazzi:2003ui}, where a remarkable match between localization results for Wilson loops and supergravity has been observed \cite{Bigazzi:2013xia},
 and to 
 backgrounds with $\mathcal{N}=1$ supersymmetry, such as the Polchinski-Strassler background \cite{Polchinski:2000uf} or its $S^4$ counterpart \cite{Bobev:2016nua}, albeit in this case no field-theory predictions are available yet. 

\subsection*{Acknowledgements}

We would like to thank R.~Borsato, A.~Dekel, K.~Pilch, D.~Sorokin, A.~Tseyt\-lin, E.~Vescovi and L.~Wulff for interesting comments and useful correspondence.
This work was supported by the Marie
Curie network GATIS of the European Union's FP7 Programme under REA Grant
Agreement No 317089, by the ERC advanced grant No 341222, by the Swedish Research Council (VR) grant
2013-4329, and by RFBR grant 15-01-99504. 

\appendix

\section{Conventions}\label{conventionsandnotations}

In this article we chose Minkowskian signature $(-++...+)$ for the background metric $G_{\mu\nu}$, while the world-sheet metric $h_{ij}$ is Euclidean with $(++)$ signature. The convention for indices used here is given by:

\begin{tabular} {l  l}
$\hat{a}$, $\hat{b}$, $\hat{c}$ = 0, 1,..., 4 & ${AdS}_{5}$ tangent space indices\\
$\hat{a'}$, $\hat{b'}$, $\hat{c'}$ = 5, 6,..., 9 & ${S}^{5}$ tangent space indices\\
$\hat{\mu}$, $\hat{\nu}$, $\hat{\rho}$, $\hat{\lambda}$.. = 0, 1,..., 9 & $AdS_{5}\times S^{5}$ tangent space indices\\
$\mu$, $\nu$, $\rho$, $\lambda$.. = 0, 1,..., 9 & $AdS_{5}\times S^{5}$ coordinate indices\\
$i$, $j$ = 0, 1 & World-sheet indices\\
I, J, K = 1, 2 & Spinor indices\\
\end{tabular}\\
The raising and lowering of the tangent space indices $\hat{\mu}$ will be done using the flat metric $\eta_{\hat{\mu}\hat{\nu}}=(-1,1,...,1)$, for the $\mu$ indices we will use the background metric tensor $G_{\mu\nu}$, while for the world-sheet indices $i,j$, we use the world-sheet metric tensor $h_{ij}$. Naturally, coordinate indices $\mu$ and tangent space indices $\hat{\mu}$ are related using the standard vierbein prescription:
\begin{align*}
{V^\mu } = {E^\mu }_{\hat \nu }{V^{\hat \nu }}\ , &&{V^{\hat \mu }} = {E_\nu }^{\hat \mu }{V^\nu }\ , && {G_{\mu \nu }} = {E_\mu }^{\hat \mu }{E_\nu }^{\hat \nu }{\eta _{\hat \mu \hat \nu }}\ .
\end{align*}

The convention used here for Dirac matrices follows the one used in \cite{Metsaev:1998it}, where the generators of the $\mathfrak{so}(4,1)$ and $\mathfrak{so}(5)$ Clifford algebras are $4\times4$ matrices $\gamma^{\hat{a}}$ and $\gamma^{\hat{a'}}$ satisfying the properties:
\begin{align}
{\gamma ^{(\hat a}}{\gamma ^{\hat b)}} &= {\eta ^{\hat a\hat b}} = \left( { -  +  +  +  + } \right) ,  & {\left( {{\gamma ^{\hat a}}} \right)^\dag } &= {\gamma ^{\hat 0}}{\gamma ^{\hat a}}{\gamma ^{\hat 0}} ,\label{cond14by4}\\
{\gamma ^{(\hat a'}}{\gamma ^{\hat b')}} &= {\eta ^{\hat a'\hat b'}} = \left( { +  +  +  +  + } \right) , & {\left( {{\gamma ^{\hat a'}}} \right)^\dag } &= {\gamma ^{\hat a'}} .
\end{align}
Just as in \cite{Metsaev:1998it}, we will choose matrices $\gamma^{\hat{a}}$ and $\gamma^{\hat{a'}}$ such that:
\begin{align}\label{conventiongamma4by4}
{\gamma ^{{{\hat a}_1}{{\hat a}_2}{{\hat a}_3}{{\hat a}_4}{{\hat a}_5}}} = i{\epsilon ^{{{\hat a}_1}{{\hat a}_2}{{\hat a}_3}{{\hat a}_4}{{\hat a}_5}}}, &&{\gamma ^{{{\hat a'}_1}{{\hat a'}_2}{{\hat a'}_3}{{\hat a'}_4}{{\hat a'}_5}}} = {\epsilon ^{{{\hat a'}_1}{{\hat a'}_2}{{\hat a'}_3}{{\hat a'}_4}\hat a'}}\ .
\end{align}

The $32\times32$ Dirac matrices used here, are constructed in terms of  $\gamma^{\hat{a}}$ and $\gamma^{\hat{a'}}$ in the following way:
\begin{align}\label{32by32Dirac}
{\Gamma ^{\hat a}} = {\gamma ^{\hat a}}\otimes\mathbbm{1}\otimes {\tau _1}, && {\Gamma ^{\hat a'}} = \mathbbm{1}\otimes {\gamma ^{\hat a'}} \otimes {\tau_2}, && \textup{C} = C \otimes C' \otimes i{\tau _2}\ ,
\end{align}
where $\mathbbm{1}$ is the $4\times4$ identity matrix, $\tau_{i}$ are the Pauli matrices, while $C$ and $C'$ are the charge conjugation matrices of the $\mathfrak{so}(4,1)$ and $\mathfrak{so}(5)$ Clifford algebras, respectively.\\

Let $\Psi$ be a 32-component spinor, here the Majorana condition takes the form of $\overline \Psi   = {\Psi ^\dag }{\Gamma ^{\hat 0}} = {\Psi ^T}\textup{C}$. In 10 dimensions, a positive chirality 32-component spinor can be decomposed in the following way: $\Psi  = \psi  \otimes \psi ' \otimes \left( {\begin{array}{*{20}{c}}
1\\
0
\end{array}} \right)= \chi \otimes \left( {\begin{array}{*{20}{c}}
1\\
0
\end{array}} \right)$, with $\chi=\psi  \otimes \psi ' $ \cite{Metsaev:1998it}. This decomposition into 16-component spinors will prove useful at several stages of the calculation. To make it more clear, let us present the following formula \cite{Metsaev:1998it}:
\begin{align}\label{exampledecomposition}
{M_{\hat \mu }}{\overline \Psi  ^I}{\Gamma ^{\hat \mu }}{\Psi ^J} = {M_{\hat a}}{\overline \chi  ^I}{\gamma ^{\hat a}}{\chi ^J} + i{M_{\hat a'}}{\overline \chi  ^I}{\gamma ^{\hat a'}}{\chi ^J}\ ,
\end{align}
here on the left hand-side ${\Gamma ^{\hat \mu }}$ corresponds to a $32\times32$ Dirac matrix as defined in equation \eqref{32by32Dirac}, while ${\Psi ^K}$ ($K=1,2$) is a 32-component $D=10$ Majorana-Weyl spinor with positive chirality. On the right hand-side, ${\chi ^K} = {\psi ^K} \otimes {{\psi '}^K}$ ($K=1,2$) is a 16-component spinor, while the $16\times16$ matrices ${\gamma ^{\hat a}}$ and ${\gamma ^{\hat a'}}$ represent ${\gamma ^{\hat a}}\otimes\mathbbm{1}$ and $\mathbbm{1}\otimes{\gamma ^{\hat a'}}$, respectively. In the main text, ${\gamma ^{\hat a}}$ and ${\gamma ^{\hat a'}}$ denote these $16\times16$ matrices unless otherwise specified. Equation \eqref{exampledecomposition} can easily be checked using \eqref{32by32Dirac} and the definitions presented above. Similar expressions, but with additional ${\Gamma ^{\hat \mu }}$ matrices are used in the process of reducing expressions with $32\times32$ matrices into lower dimensional $16\times16$ matrices.

\section{The $AdS_{5}\times S^{5}$ limit}\label{AppendexComments}

The Pilch-Warner background asymptotes to $AdS_{5}\times S^{5}$ near the boundary. To see this, take $c\rightarrow1+\frac{z^2}{2}$ for small $z$, use $dc\rightarrow z dz$ and expand equation \eqref{Einsteinm} to first order, obtaining 
\begin{align}\label{AdS5xS5metric}
ds_E^2 = \frac{{d{x^2} + d{z^2}}}{{{z^2}}} + d{\theta ^2} + {\cos ^2}\theta d{\phi ^2} + {\sin ^2}\theta d{\Omega ^2} ,
\end{align}
which is the usual metric of $AdS_{5}\times S^{5}$ with $d{\Omega ^2}$ describing the three-sphere
$$d{\Omega ^2} = \sigma _1^2 + \sigma _2^2 + \sigma _3^2\ . $$

It is important to note that for the Pilch-Warner calculation we used the classical solution $c=\sigma$, while the $AdS_{5}\times S^{5}$ computation in \cite{Drukker:2000ep,Kruczenski:2008zk} employs as classical solution $z=\sigma$. This means that the spatial world-sheet coordinates $\sigma$ of the Pilch-Warner and $AdS_{5}\times S^{5}$ computations are related by $\sigma_{\text{PW}}\rightarrow1+\frac{\sigma^{2}_{\text{AdS}}}{2}$. In order not to overload our notation, we drop the Pilch-Warner and $\text{AdS}$ labels in the $\sigma$'s, always keeping in mind the relation between the two. Having $AdS_{5}\times S^{5}$ a trivial dilaton, we see from \eqref{AdS5xS5metric} that the world-sheet metric induced by the corresponding classical solution is $ds^{2}=\frac{1}{\sigma^{2}}\left(d\tau^{2}+d\sigma^{2}\right)$.\\
\indent For completeness, we now apply the limiting procedure to the bosonic and fermionic operators presented in sections \ref{bosonsPW} and \ref{fermionsPW}. To obtain the appropriate $AdS_{5}\times S^{5}$ operators it is necessary to simultaneously make the substitutions $\sigma\rightarrow1+\frac{\sigma^{2}}{2}$ and $\partial_{\sigma}\rightarrow\frac{1}{\sigma}\partial_{\sigma}$, and then expand to first order in $\sigma\rightarrow0$. For the bosonic operators in  \eqref{HIs}-\eqref{HIII.}, this results in
\begin{align}\label{PWtoAdSBosons}
{\mathcal{K}_\mathbf{x}}&\rightarrow - \partial _\tau ^2 - \partial _\sigma ^2 + \frac{2}{\sigma }{\partial _\sigma }\ ,   & N_\mathbf{x}=3\nonumber\\
{\mathcal{K}_\phi}&\rightarrow - \partial _\tau ^2- \partial _\sigma ^2 + \frac{2}{\sigma }{\partial _\sigma } - \frac{2}{{{\sigma ^2}}}\ ,     & N_\phi=1\\
{\mathcal{K}^{\pm}_\mathbf{y}}&\rightarrow - \partial _\tau ^2- \partial _\sigma ^2 + \frac{2}{\sigma }{\partial _\sigma } - \frac{2}{{{\sigma ^2}}}\ ,     & N^{\pm}_\mathbf{y}=2\nonumber
\end{align}
where the linear time derivative in $\mathcal{K}^{\pm}_\mathbf{y}$ does not contribute to first order in $1/\sigma$, just as expected as $AdS_{5}\times S^{5}$ has no $B$-field. As we will see in appendix \ref{TheAds5xS5CaseBosons}, the above bosonic operators are related to the ones found in \cite{Drukker:2000ep,Kruczenski:2008zk}.\\
\indent To take the $AdS_{5}\times S^{5}$ limit of the fermionic operator \eqref{16by16fermionicoperator}, we will perform the same limiting procedure, while ignoring the terms with $c_{(1)}^{\rm{RR}}$, $c_{(3)}^{\rm{RR}}$ and $c_{(3)}^{\rm{NSNS}}$, as the only R-R flux in $AdS_{5}\times S^{5}$ is the five-form and there is no NS-NS three-form. After the required substitutions and expansion, the final result is
\begin{align}\label{fermionsads5xs5}
L_F^{\left( 2 \right)}\rightarrow2\sqrt h \overline \chi  \left[ {\sigma {\gamma ^{\hat 1}}{\partial _\tau } + \sigma {\gamma ^{\hat 4}}{\partial _\sigma } - \frac{1}{2}{\gamma ^{\hat 4}} - {\gamma ^{\hat 1\hat 4}}} \right]\chi \ ,
\end{align}
where $\chi$ is a 16-component spinor and $\gamma^{\hat a}$ are the $16\times16$ matrices described in appendix \ref{conventionsandnotations}. In equation \eqref{fermionsads5xs5}, the first two terms correspond to the kinetic terms, while the third and the fourth come from contributions of the spin-connection and the R-R five-form, respectively.

It is interesting to note that the above fermionic operator differs slightly from the one in \cite{Drukker:2000ep,Kruczenski:2008zk} because we chose our classical solution to be $x^{1}=\tau$ in order to have a world-sheet metric with Euclidean signature. Had we considered a classical solution $x^{0}=\tau$, the resulting fermionic operator would be exactly the same as in \cite{Drukker:2000ep,Kruczenski:2008zk}, but would have a Minkowskian world-sheet signature.

\section{String partition function in $AdS_{5}\times S^{5}$}\label{TheAds5xS5Case}

The calculation of the semiclassical partition function for the straight string in the $AdS_{5}\times S^{5}$ background was first done in \cite{Drukker:2000ep}. The straight Wilson line in $AdS_{5}\times S^{5}$ has trivial expectation value, not renormalized by quantum corrections. A particularly simple, symmetry-based argument for cancellation of the one-loop partition function for the straight line in $AdS_5\times S^5$ is given in the appendix~B of \cite{Buchbinder:2014nia}. Here we illustrate how the cancellation of the one-loop quantum corrections is reproduced within the formalism that we use in the main text for the Pilch-Warner background.\\
\indent First, we will present the corresponding contributions from bosons and fermions, whose resulting operators have a structure similar to the $\mathcal{K}_\mathbf{x}$, $\mathcal{K}_\phi$ and $\mathcal{D}_{0}$ Pilch-Warner operators. Then, we will see how the corresponding determinants produce the expected result for the $AdS_{5}\times S^{5}$ case. Due to the much simpler field content of $AdS_{5}\times S^{5}$, there is no need for numerics as equations can be solved analytically, making this an ideal test ground for the consistency check of the formalism that we use.

\subsection{Bosonic Fluctuations}\label{TheAds5xS5CaseBosons}

As in \cite{Kruczenski:2008zk}, we will assume cancellation between ghosts and the bosonic fluctuations along the longitudinal modes $\zeta^{\hat{1}}$ and $\zeta^{\hat{4}}$, since their actions are identical\footnote{In our convention the classical solution is oriented along $x^{1}=\tau$ and not $x^{0}$ as in \cite{Drukker:2000ep,Kruczenski:2008zk}.}. Thus, having $AdS_{5}\times S^{5}$ a vanishing Fradkin-Tseytlin term, the bosonic contribution to the semiclassical partition function will consist exclusively of the quadratic transverse fluctuations.  As shown in \cite{Drukker:2000ep,Kruczenski:2008zk}, out of the 8 transverse modes; 5 will be massless modes (which come from $S^{5}$ fluctuations), while 3 will have mass squared equal to 2 (which correspond to the remaining $AdS_{5}$ transverse modes). The contribution of these fluctuations is given by the action \cite{Drukker:2000ep,Kruczenski:2008zk}
\begin{align}\label{ads5xs5bosons}
{S_{2B}} = \frac{{\sqrt \lambda  }}{{4\pi }} \! \int \!{\frac{d\tau d\sigma}{{{\sigma ^2}}}\left[\ \ \ {\sum\limits_{\mathclap{\hat a,\hat b\, \in \left\{ {0,2,3} \right\}}} {{\eta _{\hat a\hat b}}{\zeta ^{\hat a}}\!\left( {{\sigma ^2}( - \partial _\tau ^2 - \partial _\sigma ^2) + 2} \right)\!{\zeta ^{\hat b}} + \sum\limits_{\mathclap{\hat a = 5}}^9 {{\zeta ^{\hat a}}\!\left( {{\sigma ^2}( - \partial _\tau ^2 - \partial _\sigma ^2)} \right)\!{\zeta ^{\hat a}}} } } \right]} .
\end{align}
Comparing with the standard normalization for bosons $\int {\sqrt h \zeta \zeta d\tau d\sigma }$, we see that the bosonic operators obtained from \eqref{ads5xs5bosons} will have a factor of $\sigma^{2}$ in front of $\partial^{2}_{\tau}$. To make the calculation simpler, we will remove this factor by performing a field redefinition analogous to the one done for bosons in section \ref{bosonsPW}
\begin{align}\label{rescalingbosons}
{\zeta ^{\hat a}} \to \frac{1}{\sigma }{\xi ^{\hat a}} .
\end{align}
Naturally, this field redefinition will modify the measure of the bosonic path integral, but as we will see later, this will be compensated by a similar factor from a fermionic field redefinition. Doing this redefinition and using partial integration, the bosonic action can be written in the following way
\begin{align}
{S_{B}^{(2)}} = \frac{{\sqrt \lambda  }}{{4\pi }}\int\! \frac{d\tau d\sigma}{{{\sigma ^2}}} & \left[\ \ \ \ \sum\limits_{\mathclap{\hat a,\hat b\, \in \left\{ {0,2,3} \right\}}} {{\eta _{\hat a\hat b}}{\xi ^{\hat a}}\left( {-\partial _\tau ^2 - \partial _\sigma ^2 + \frac{2}{\sigma }{\partial _\sigma }} \right){\xi ^{\hat b}}}\right.\nonumber\\
&\left.  \quad\quad\quad+ \sum\limits_{\hat a = 5}^9 {{\xi ^{\hat a}}\left( {-\partial _\tau ^2 - \partial _\sigma ^2 + \frac{2}{\sigma }{\partial _\sigma } - \frac{2}{{{\sigma ^2}}}} \right){\xi ^{\hat a}}}   \right] .
\end{align}
From this expression it is clear that the bosonic contribution to the partition function is given by the determinant of 2 differential operators:
\begin{align}\label{BosonsZ}
{Z_{\text{Bosons}}} \propto {\det^{- 3/2}}\left( {-\partial _\tau ^2 - \partial _\sigma ^2 + \frac{2}{\sigma }{\partial _\sigma }} \right) {\det^{- 5/2}}\left( {-\partial _\tau ^2 - \partial _\sigma ^2 + \frac{2}{\sigma }{\partial _\sigma } - \frac{2}{{{\sigma ^2}}}} \right) ,
\end{align}
which are naturally the $AdS_{5}\times S^{5}$ limits of the bosonic Pilch-Warner operators (recall equations \ref{PWtoAdSBosons}).

\subsection{Fermionic Fluctuations}
We will take as starting point the $AdS_{5}\times S^{5}$ limit of the fermionic Pilch-Warner operator (see equation \eqref{fermionsads5xs5}). By using the convenient representation of Dirac matrices of equation \eqref{choiceofgammas}, we can write equation \eqref{fermionsads5xs5} more explicitly in the following way\footnote{Alternatively, one can use the Minkoswkian signature operator of \cite{Drukker:2000ep,Kruczenski:2008zk}, in which case one would need a slightly different choice of representation for the Dirac matrices.}
\begin{align}
{L_{F}^{(2)}} =  2\sqrt h \ \sum\limits_{j = 1}^8 {\left( {\begin{array}{*{20}{c}}
{{{\bar \chi }_j}}&{{{\bar \chi }_{j + 8}}}
\end{array}} \right)} \left( {\begin{array}{*{20}{c}}
{ - \sigma {\partial _\tau }}&{  \sigma {\partial _\sigma } + \frac{1}{2}}\\
{  \sigma {\partial _\sigma } - \frac{3}{2}}&{\sigma {\partial _\tau }}
\end{array}} \right)\left( {\begin{array}{*{20}{c}}
{{\chi _j}}\\
{{\chi _{j + 8}}}
\end{array}} \right)\ .
\end{align}
This means that the fermionic operator $\mathcal{K}_{F}$, normalized as $L_{2F}=2\sqrt{h}\ \bar{\chi}\mathcal{K}_{F}\chi$, can be seen as a $16\times16$ block matrix composed of 8 identical $2\times2$ blocks. In order to cancel the $\sigma$ in front of the $\partial_{\tau}$ derivatives, we perform the following field redefinition
\begin{align}
{\chi _i} \to \frac{1}{{\sqrt \sigma  }}\ {\psi _i}\ .
\end{align}
At the level of the partition function, this scaling of the 16 components $\chi_{i}$ will produce a factor in the measure of the path integral which precisely cancels the one produced by the scaling of the 8 transverse bosonic fluctuations (recall equation \eqref{rescalingbosons}). After performing this rescaling and a relabelling of the indices, the fermionic Lagrangian can be written as
\begin{align}
{L_{F}^{(2)}} =  2\sqrt h \sum\limits_{i = 1}^8 {\left( {\begin{array}{*{20}{c}}
{{{\bar \psi }_{2i - 1}}}&{{{\bar \psi}_{2i}}}
\end{array}} \right)} \left( {\begin{array}{*{20}{c}}
{{-\partial _\tau }}&{ {\partial _\sigma } }\\
{ {\partial _\sigma -\frac{2}{\sigma} }}&{  {\partial _\tau }}
\end{array}} \right)\left( {\begin{array}{*{20}{c}}
{{\psi _{2i - 1}}}\\
{{\psi _{2i}}}
\end{array}} \right)\ .
\end{align}
Consequently, the contribution of fermions to the semiclassical partition function can be written in terms of the determinant of a $2\times2$ operator
\begin{align}\label{Zfermions1}
{Z_{\text{Fermions}}} \propto {\det ^4}\left( {\begin{array}{*{20}{c}}
{{-\partial _\tau }}&{  {\partial _\sigma }}\\
{  {\partial _\sigma }-\frac{2}{\sigma}}&{ {\partial _\tau }}
\end{array}} \right)\ .
\end{align}
As is usually done with fermions, instead of evaluating the determinant of the operator in \eqref{Zfermions1}, we consider the square of this operator
\begin{align}
{Z_{\text{Fermions}}}& \propto {\det ^2}{\left( {\begin{array}{*{20}{c}}
{\partial _\tau ^2 + \left( { {\partial _\sigma } } \right)\left( {  {\partial _\sigma }-\frac{2}{\sigma}} \right)}&0\\
0&{\partial _\tau ^2 + \left( { {\partial _\sigma }-\frac{2}{\sigma}} \right)\left( { {\partial _\sigma } } \right)}
\end{array}} \right)},\nonumber\\
& \propto{\det ^2}\left( {-\partial _\tau ^2 - \partial _\sigma ^2 + \frac{2}{\sigma }{\partial _\sigma } - \frac{2}{{{\sigma ^2}}}} \right) {\det ^2}\left( {-\partial _\tau ^2 - \partial _\sigma ^2 + \frac{2}{\sigma }{\partial _\sigma }} \right)\ .\label{FermionsZ}
\end{align}
Note that after squaring, the operators found above are the same as the ones present in the bosonic contribution \eqref{BosonsZ}. This intricate relation between bosons and fermions is similar to the one observed for the Pilch-Warner operators of equation \eqref{D4-K1-K2}.

\subsection{The Semiclassical Partition Function}

Combining the contributions of the bosonic and fermionic operators of equations \eqref{BosonsZ} and \eqref{FermionsZ}, we have that the semiclassical partition function is given by\footnote{We assume that the spectrum of operators is the same when appearing as bosons and fermions. This is a consequence of the choice of boundary conditions.}
\begin{align}
Z &\propto {\det ^{1/2}}\left( {-\partial _\tau ^2 - \partial _\sigma ^2 + \frac{2}{\sigma }{\partial _\sigma }} \right){\det ^{ - 1/2}}\left( {-\partial _\tau ^2 - \partial _\sigma ^2 + \frac{2}{\sigma }{\partial _\sigma } - \frac{2}{{{\sigma ^2}}}} \right)\nonumber\\
&\propto \text{Exp}\left[ {\frac{1}{2}\frac{\rm{L}}{{2\pi }}\int {\text{tr}\ln ({\omega ^2} +\mathcal{H}_{1}) - \text{tr}\ln ({\omega ^2} +\mathcal{H}_{2})d\omega } } \right]\ ,\label{Zgettingclose}
\end{align}
where we first Fourier expanded in $\tau$, took the continuum limit in $\omega$ and used the following definitions for the operators $\mathcal{H}_{1,2}$
\begin{align}\label{h1h2ads}
{\mathcal{H}_1} =  - \partial _\sigma ^2 + \frac{2}{\sigma }{\partial _\sigma }\ , && {\mathcal{H}_2} =  - \partial _\sigma ^2 + \frac{2}{\sigma }{\partial _\sigma } - \frac{2}{{{\sigma ^2}}}\ ,
\end{align}
which have the same behaviour at large $\sigma$: ${\mathcal{H}_0} =  - \partial _\sigma ^2$.\\

In order to evaluate $Z$ in equation \eqref{Zgettingclose}, we will consider the following spectral problems
\begin{align}\label{AdS5xS5thethreeH}
{{\mathcal{H}}_i}{\psi _i} = p_i^2{\psi _i}\quad \text{with}\quad i\in\{0,1,2\} .
\end{align}
First, we will present an explicit computation using the phaseshift method and later we put forward an argument based on the isospectral structure of $\mathcal{H}_{1,2}$ and the choice of boundary conditions. Since the operators in  \eqref{h1h2ads} are relatively simple, there is no major gain in rewriting the spectral problems \eqref{AdS5xS5thethreeH} in flat space. Moreover, having no linear time derivatives in the operators, it is much simpler to treat them in their $1\times1$ form without recurring to a $2\times2$ representation, as was done for the PW case.\\

As discussed in section \ref{spectralproblem}, the solution to the second order differential equation \eqref{AdS5xS5thethreeH} will be a superposition of 2 solutions with different power-like behaviour as $\sigma\rightarrow0$. In general, we pick the solution with the highest power of $\sigma$ as this will provide an adequate normalization condition. As in the Pilch-Warner case, string fluctuations will be thought as being described by the Schr\"odinger problem of a particle oscillating in between two infinite walls
\begin{align}\label{boundaryAdS5xS5}
\psi_{i}(\sigma=0)=0\ ,&& \psi_{i}(\sigma=R)=0\ , && \forall i\in\{0,1,2\}
\end{align}
where the spectrum is discrete since the equation on the right can be thought of as a quantization condition.\\

For the asymptotic operator $\mathcal{H}_{0}$, we see that the eigenfunction $\psi_{0}$ has an oscillatory behaviour
$${\psi _0} = {A_0}\sin \left( {{p_0}\ \sigma } \right) ,$$
where $A_0$ denotes the amplitude of oscillation. For the non-asymptotic operators $\mathcal{H}_{1}$ and $\mathcal{H}_{2}$, we proceed as in the Pilch-Warner background and assume that for large $\sigma$ the eigenfunctions are of the form
\begin{align}\label{ansatznonasymptotic}
\lim_{\sigma\rightarrow\infty}{\psi _i} \approx {A_i}\sin \left( {{p_i}\sigma  + {\delta _i}\left( p_{i}  \right)} \right) .
\end{align}
Due to the phaseshift $\delta_{i}$, the quantization condition for the asymptotic $\mathcal{H}_{0}$ and non-asymptotic operators $\mathcal{H}_{i}$ $\left( i\in\{1,2\}\right)$ will be given by
\begin{align}
{p_{0}} R = \pi n\ ,&& {p_{i}} R + {\delta _i}\left( {{p_{i}}} \right) = \pi n\ ,\nonumber
\end{align}
respectively. Naturally, this implies the following density of states for $\mathcal{H}_0$ and $\mathcal{H}_{i}$ ($i\in\{1,2\}$)
\begin{align}\label{quantizationAdS5xS5}
\frac{{d n}}{{d {p_{0}}}} = \frac{R }{\pi}\ ,&& \frac{{d n}}{{d {p_{i}}}} = \frac{R }{\pi} + \frac{1}{\pi }{\delta _i}'\left( {{p_{i}}} \right)\ .
\end{align}
By adding zero in a convenient way and using equations \eqref{AdS5xS5thethreeH} and \eqref{quantizationAdS5xS5}, we can rewrite the partition function as
\begin{align}
Z &\propto {\rm{Exp}}\left[ \frac{1}{2}\frac{\rm{L}}{{2\pi }}\int \left( {\text{tr}\ln ({\omega ^2} + {{\mathcal{H}}_1}) - \text{tr}\ln ({\omega ^2} + {{\mathcal{H}}_0})} \right) \right.\nonumber\\
&\left. \quad\quad\quad\quad\quad - \left( {\text{tr}\ln ({\omega ^2} + {{\mathcal{H}}_2}) - \text{tr}\ln ({\omega ^2} + {{\mathcal{H}}_0})} \right)d\omega  \right]\ ,\nonumber\\
& \propto {\rm{Exp}}\left[ \frac{1}{2}\frac{\rm{L}}{{2\pi }}\int \left[ \left( {\int {\ln ({\omega ^2} + p_1^2)\frac{{dn}}{{d{p_1}}}d{p_1}}  - \int {\ln ({\omega ^2} + p_0^2)\frac{{dn}}{{d{p_0}}}d{p_0}} } \right)\right.\right.\nonumber\\
& \left.\left. \quad\quad\quad\quad\quad - \left( {\int {\ln ({\omega ^2} + p_2^2)\frac{{dn}}{{d{p_2}}}d{p_2}}  - \int {\ln ({\omega ^2} + p_0^2)\frac{{dn}}{{d{p_0}}}d{p_0}} } \right) \right]d\omega   \right]{\kern 1pt}, \nonumber\\
& \propto {\rm{Exp}}\left[ {\frac{1}{{2\pi }}\frac{\rm{L}}{{2\pi }}\int {\int {\ln ({\omega ^2} + {p^2})\left[ {\delta {'_1}\left( p \right) - \delta {'_2}\left( p \right)} \right] dp}\ d\omega } } \right]{\kern 1pt} .\label{AdS5xS5ZSecondtolast}
\end{align}
Thus, to evaluate $Z$ all that is left is to evaluate the phaseshifts $\delta_i$. In order to do this, we first find the solutions to equations \eqref{AdS5xS5thethreeH} subject to the chosen boundary conditions 
\begin{align}\label{eigenfuinctionsAdS5xS5}
{\psi _1} &= {A_1}\left[ {\sin \left( {{p_1}\sigma } \right) + {p_1}\sigma \sin \left( {{p_1}\sigma  - \frac{\pi }{2}} \right)} \right]\ ,\nonumber\\
{\psi _2} &= {A_2}\ \sigma \sin \left( {{p_2}\sigma } \right)\ .
\end{align}
By comparing the large $\sigma$ behaviour of these solutions with equation \eqref{ansatznonasymptotic}, we see that $\delta_{1}=-\frac{\pi}{2}$ and $\delta_{2}=0$. Replacing these results in \eqref{AdS5xS5ZSecondtolast}, we obtain the well-known result for the semiclassical partition function
$$Z\propto 1\ .$$
Alternatively, we could have arrived at this result by considering the structure of the operators $\mathcal{H}_{1,2}$
\begin{align}
{\mathcal{H}_1} = L{L^\dagger}\ , && {\mathcal{H}_2} =  {L^\dagger} L\ ,\nonumber
\end{align}
where $L=\partial_{\sigma}-\frac{2}{\sigma}$ and ${L^\dagger}=-\partial_{\sigma}$, with the latter being conjugate operators with respect to the scalar product $\langle {\psi _1} | \psi _{2}\rangle  = \iint {\frac{{d\tau d\sigma }}{{{\sigma ^2}}}}{\psi _1}{\psi _2}$.\\

Notice that the pair of operators $\mathcal{H}_{1,2}$ have the same structure as the PW operators in \eqref{factorized}. As explained in equations \eqref{factorized}-\eqref{K2=K1}, operators of this type have the  same spectrum (up to zero modes of $L$ and ${L^{\dagger}}$)\footnote{Note that the zero modes of $L$ and ${L^\dagger}$,  $\psi\propto\sigma^{2}$ and $\psi=Const$, respectively, do not have oscillatory behaviour at large $\sigma$ and are therefore excluded in the phaseshift computation.} provided that the map between the eigenfunctions $\psi_{1}\propto L \psi_{2}$ and $\psi_{2}\propto{L^\dagger}\psi_{1}$ is compatible with the choice of boundary conditions. Indeed, by explicit computation it can be checked that the eigenfunctions satisfying our choice of boundary conditions (see \eqref{eigenfuinctionsAdS5xS5}) are mapped into each other by $L$ and ${L^{\dagger}}$.

\section{Second order differential equations}\label{sec:2ndODEs}
The Dirac eigenvalue problem \eqref{sch2} can be decoupled into two second order differential equations of the form $O_{n}\phi_{n}=0$ with $n\in\{1,2\}$ (one for each component of the eigenvector). 
Here we will write down the explicit equations for our operators $\hat{\mathcal{H}}_{B,F}$:
\begin{eqnarray} \label{2ndODEcommon}
O_{B1, F2} & = &
		-(\sigma ^2-1) A^2 \partial_{\sigma}^2
		+ 2 A (2-3 \sigma  A) \partial_{\sigma} \nonumber\\
	  & &	+  1 - E ^2 + (1+E) V(\sigma) + U_{B1, F2}(\sigma),\\
O_{B2, F1} & = &
		-(\sigma ^2-1) A^2 \partial_{\sigma}^2
		+ \left(2 A (2-3 \sigma  A) +\frac{\left(\sigma ^2-1\right) A^2 V'(\sigma )}{1-E+V(\sigma )}\right) \partial_{\sigma} \nonumber\\
	  & &	+  1 - E ^2 + (1+E) V(\sigma) + U_{B2, F1}(\sigma),	  
\end{eqnarray}
where the different $U(\sigma)$ are:
\begin{eqnarray*}
U_{B1}(\sigma) & = & \frac{A}{4 \sigma ^2 (\sigma ^2-1)} \left(\left(-24 \sigma ^4+6 \sigma ^2+3\right) A+16 \sigma ^3\right), \\
U_{F1}(\sigma) & = & \frac{A }{4 \sigma ^2 \left(\sigma ^2-1\right)}\left(\left(-24 \sigma ^4+6 \sigma ^2+3\right) A+16 \sigma ^3\right)\\
		& &  +\frac{\left(\left(4 \sigma ^2-1\right) A-4 \sigma \right) A V'(\sigma)}{2 \sigma  (1-E+V(\sigma))},\\	
U_{B2}(\sigma) & = & \frac{A}{4 \sigma ^2 \left(\sigma ^2-1\right)} \left(\left(-16 \sigma ^4+2 \sigma ^2-1\right) A+ 16 \sigma ^3+8\sigma \right)\\
		& & +\frac{\left(2 \sigma ^2+1\right) A^2 V'(\sigma)}{2 \sigma  (1-E+V(\sigma))},\\
U_{F2}(\sigma) & = & \frac{A}{4 \sigma ^2 (\sigma ^2-1)} \left(\left(-16 \sigma ^4+2 \sigma ^2-1\right)A +16 \sigma ^3+8\sigma\right).
\end{eqnarray*}
$E$ is the eigenvalue \eqref{eigenE} and $V(\sigma) = A(\sigma) / \sigma$.

\section{WKB expansion of phaseshifts}
\label{large-p}

The WKB approximation applies to linear differential equations with a small parameter multiplied to the highest derivative term.
In our case, we will use the decoupled second order equations, shown in appendix \ref{sec:2ndODEs},
and we will take the momentum $p$ to be large (or $1/p $ to be small). 

Our WKB ansatz is written as
\begin{equation}  \label{WKBansatz}
 \phi(\sigma) = e^{i S(\sigma)}, \quad S(\sigma)=p \sum_{i=0}^{n}  p^{-i} S_{i}(\sigma) \qquad (p \rightarrow \infty).
\end{equation}
Expansion in powers of $p$ reduces each differential equation to a set of coupled algebraic equations for $S_i'(\sigma)$ that can be solved recursively. 
Then,
\begin{equation}
 S_i(\sigma) = \int_{1}^{\sigma} S_i'(x) \,dx.
\end{equation}

Since the asymptotic solutions of our differential equations are plane waves \eqref{boundaryCs}, 
there are actually two sets of solutions for $S_i'$, whose imaginary parts are the same but their real parts differ by a sign.
Thence the exponentials combine to sine (or cosine).
The imaginary part of $S$ gives the amplitude, which is irrelevant for the computation of the determinant,
and the real part is related to the phaseshift by:
\begin{equation} \label{relationSphaseshift}
 \text{Re} ( S(\sigma) )= p \sigma + \delta(p) \qquad (\sigma \rightarrow \infty).
\end{equation}
The explicit WKB solutions for the first component equations are shown in the next subsection \ref{sec:WKBsolutions}. In subsection \ref{sec:WKBCancellation}, we will show  the cancellation of UV divergences for $\Delta$ in \eqref{deltaphases}. 
Then, in \ref{sec:E3}, we will use the WKB method to compute $\delta(p)$ up to order $\mathcal{O}(1/p)$.

\subsection{WKB solutions} \label{sec:WKBsolutions}

One set of the solutions for the WKB ansatz \eqref{WKBansatz}, for the first component equations, i.e. \eqref{2ndODEcommon} with $U_{B1}$ and $U_{F1}$, with positive and negative energy (subindexes $B$, $F$, $+$, and $-$, respectively) is given by:
\begin{eqnarray*}  \label{WKBsolutions}
  S_{0, B, F,\pm}' & = & \frac{2}{3 A \sqrt{\sigma ^2-1} }  , \\     
  S_{1, B, F,\pm}' & = & \mp \frac{1}{2 \sigma  \sqrt{\sigma ^2-1}} 
   + i \frac{2-3 \sigma  A}{2 A(1- \sigma^2)}, 
  \\        
  S_{2, B, \pm}' & = & \frac{3 \left(\left(3 \sigma ^4-\sigma ^2-2\right) A^2+4 \sigma ( A - \sigma )\right)}{16 A \sigma ^2 \left(\sigma ^2-1\right)^{3/2}} 
   \mp i \,\frac{3 \left(\left(\sigma ^2+1\right) A-2 \sigma \right)}{8 \sigma ^2 \left(\sigma ^2-1\right)},
  \\  
  S_{2, F,\pm}' & = & \frac{3 \left(\left(3 \sigma ^4-\sigma ^2-2\right) A^2+4 \sigma ( A - \sigma )\right)}{16 A \sigma ^2 \left(\sigma ^2-1\right)^{3/2}} 
   \pm i \,\frac{3 \left(\left(\sigma ^2+1\right) A-2 \sigma \right)}{8 \sigma ^2 \left(\sigma ^2-1\right)},
  \\    
  S_{3,B,\pm}' & = & \pm
 \frac{9 \left(-8 \left(\sigma ^2+ 1 \right)\sigma A+\left(9 \sigma ^4+\sigma ^2+2\right) A^2-4 \left(\sigma ^2-2\right) \sigma ^2\right)}{64 \sigma ^3 \left(\sigma ^2-1\right)^{3/2}}\\ 
 		& & + i \, \frac{9 \left(-2 \sigma \left(3 \sigma ^2+2\right)  A+\left(6 \sigma ^4+\sigma ^2+1\right) A^2+4 \sigma^2\right)}{32  \sigma ^3 \left(\sigma ^2-1\right)} ,   
		\\     
  S_{3,F,\pm}' & = & \pm
 \frac{9 \left(-8 \left(\sigma ^2-2\right) \sigma  A+\left(\sigma ^4-3 \sigma ^2-10\right) A^2-4 \left(\sigma ^2-2\right) \sigma ^2\right)}{64 \sigma ^3 \left(\sigma ^2-1\right)^{3/2}}  \\ 
  		& & + i \, \frac{9 \left(-2 \sigma \left(\sigma ^2+4\right)  A+\left(6 \sigma ^4+5\sigma ^2+5\right) A^2-4 \sigma^2\right)}{32  \sigma ^3 \left(\sigma ^2-1\right)} .   
\end{eqnarray*}
The other set of solutions is obtained by changing the signs of the real part of the solutions above.

Higher order WKB terms can be computed in the corresponding \emph{Mathematica} notebook\footnote{
See the online repository \url{github.com/yixinyi/PhaseShiftMethod}.}.

\subsection{Cancellation of divergences}\label{sec:WKBCancellation}
Given the WKB solutions in section \ref{sec:WKBsolutions}, 
we observe that the bosonic and fermionic modes are the same for $S_{i}'$ with $i=0,1,2$ (up to signs for particles and holes),
which implies exact cancellation of the UV divergences in \eqref{deltaphases}. 
Furthermore, after integration, the phaseshift contribution of $S_3'$ terms cancels too. 
This means that the first non-zero order is $\mathcal{O}(p^{-3})$, which comes from the $S_4'$ terms. 

Actually, particles and their respective holes differ by a sign in even WKB orders, namely in odd powers of $1/p$. 
This means that the next-to-leading order correction comes from $S_6'$.

Using \eqref{relationSphaseshift}, the phaseshift difference at large $p$ is:
\begin{eqnarray} \label{differencePhaseshiftsWKB}
\delta_F ^+ + \delta_F ^- -\delta_B ^+ - \delta_B^- &=& \frac{9 \pi ^2 \left(256-96 \pi +45 \pi ^2\right)}{2048 \, p^3}  \nonumber\\
	& & -\frac{81 \pi ^3 \left(860160-702848 \pi -3880800 \pi ^2+1245825 \pi ^3\right)}{18350080 \, p^5}  + \ldots \nonumber\\
	& = & 17.2856 \, p^{-3} + 139.805 \, p^{-5} + \ldots 
\end{eqnarray}

\subsection{Large momentum expansion for phaseshifts} \label{sec:E3}
The large-$p$ behaviour for $\delta$ is
\begin{equation}  \label{largepPhaseshift}
 \delta(p) = p \, \delta_{0} + \delta_{1} + \frac{1}{p} \,\delta_{2} + \ldots \qquad (p\rightarrow \infty).
\end{equation}
Let us compute some numeric coefficients $\delta_{i}$, which will be used to test the numeric results.

For the leading order, we absorb the linear $\sigma$ term in \eqref{relationSphaseshift} into the integration by using the identity $\sigma = \int_1^{\sigma} dx + 1$.
This regularizes the integrand at infinity.
We will also compute the next-to-leading order term in the limit $\sigma \rightarrow \infty$, by splitting the integration domain as shown below:
\begin{eqnarray} \label{phaseshift0}
  \delta_0
	   & = &  \int_1^{\sigma} (S_0'(x)-1 ) \, dx  -  1, \nonumber \\
	   & = &  \int_1^{\infty} (S_0'(x)-1 ) \, dx  -  1  - \int_{\sigma}^{\infty} (S_0'(x)-1 ) \, dx, \nonumber\\
	   & = &  \int_1^{\infty} \left(\frac{2}{3 A(x) \sqrt{x ^2-1} } - 1 \right) \, dx - 1 - \int_{\sigma}^{\infty} \left(\dfrac{3}{10} x^{-2} + O(x^{-3}) \right ) \, dx, \nonumber \\
	   & \approx & -0.384 - \dfrac{3}{10} \sigma^{-1} + \mathcal{O}(\sigma^{-2}),
\end{eqnarray}
where we used the expansion 
\begin{equation} 
  S_0'(\sigma) \approx 1 + \frac{3}{10} \sigma^{-2} + \mathcal{O}(\sigma^{-3})
  \qquad (\sigma \rightarrow \infty).
\end{equation}
The subleading term in \eqref{phaseshift0} will be used to estimate the error of our numeric algorithm, as explained in appendix \ref{sec:numericError}.

The higher order terms in large $p$ are obtained by straightforward integration:
\begin{equation*}
 \delta_i = \int_1^{\infty} \text{Re}\left(S_i'\left(x\right) \right) \, dx, \quad i=1,2.
\end{equation*}
The final phaseshift expansion, common to both bosonic and fermionic\\ modes, is:
\begin{equation} \label{WKBphaseshift}
 \delta_{B,F,\pm}(p) \approx -0.384 \, p \mp 0.785-\frac{1.32}{p} + \mathcal{O}(p^{-2}).
\end{equation}

\section{Numeric error estimate} \label{sec:numericError}

Our numeric algorithm\footnote{
The corresponding \emph{Mathematica} code can be found in the online repository \url{github.com/yixinyi/PhaseShiftMethod}.}
consists of three main parts:
\begin{enumerate}
 \item Solving numerically the differential equations \eqref{sch2} with boundary conditions \eqref{sigmato1solutions} at $\varepsilon=\sigma-1$, 
  for a region of order $\sigma_{\text{max}}$ that we chose to be the interval
  $[\sigma_{\text{max}}-3\lambda,\sigma_{\text{max}}]$, where $\lambda=\frac{2\pi}{p}$ denotes the wavelength.
 \item Fitting of the numerical solutions to cosine in order to find the phaseshifts. 
 \item Numerical integration of \eqref{deltaphases} over a finite range $[p_{\text{min}}, \; p_{\text{max}}]$.
\end{enumerate}
The numeric parameters used are $\varepsilon = 10^{-6}$ and $\sigma_{\text{max}} = 1000$ (smaller $\varepsilon$ does not improve the result, and larger $\sigma_{\text{max}}$ takes much longer time). We integrate from $p_{\text{min}} = 0.1$ to $p_{\text{max}} = 50$ in steps of $\delta p = 0.1$, hence we have $\text{N}=500$ points.

The numeric integration error can be estimated by approximating the integration by a sum, namely for
\begin{equation}
 \Delta = \sum_{i=1}^{\text N} f_i \, \delta p, 
\end{equation}
the standard error propagation formula gives
\begin{equation}
 \text{error}(\Delta) = \delta p \,  \sqrt{\sum_{i=1}^{\text N}  \text{error}(f_i)^2},
\end{equation}
where 
\begin{equation}
 f_i = \dfrac{16 p_i}{9 \pi\sqrt{\frac{4}{9} p_i^2+1}} (\delta_{F,i}^+ +\delta_{F,i}^- -\delta_{B,i}^+ - \delta_{B,i}^-), \quad p_i = i \, \delta p. 
\end{equation}

Let us estimate the error of the phaseshifts. 
We consider the finiteness of $\sigma_{\text{max}}$ as the dominant source,
and it is estimated from the finite-$\sigma$ correction in \eqref{phaseshift0}.
Therefore\footnote{Higher order $S_n'(\sigma) \sim \mathcal{O}(\sigma^{-2})$ as well, but they are subleading in large $p$.},
\begin{equation}
 \text{error}(\delta) =  - \dfrac{3\, p}{10 \,\sigma_{\text{max}}}.
\end{equation}
Though this estimate is valid for large $p$, it is reasonable to assume it applies for the whole integration range, because at small $p$ phaseshifts are suppressed by the factor in the integrand that multiplies the phaseshifts. 

Attributing the same error to all the phaseshifts (giving an additional factor of $2=\sqrt{4}$), the integrand error is 
\begin{equation}
\text{error}(f_i) = \dfrac{16 p_{i}}{9 \pi\sqrt{\frac{4}{9} p_{i}^2+1}} 2 \, \left(- \frac{3 \, p_i}{10 \,\sigma_{\text{max}}}\right).
\end{equation}

Putting all the numbers together, we have that
\begin{equation}
 \text{error} (\Delta) = \pm 0.03.
\end{equation}
Moreover, we can use the WKB approximation \eqref{differencePhaseshiftsWKB} to estimate the tail $(p_{\text{max}}, \infty)$ contribution:
\begin{equation}
 \text{error}(\Delta)_{p_{\text{max}}}= \int_{p_{\text{max}}}^\infty \dfrac{16 p}{9 \pi\sqrt{\frac{4}{9} p^2+1}} \dfrac{17.2856}{p^3} dp \approx 0.003,
\end{equation}
which is much smaller, hence the total error is the numeric error:
\begin{equation}
  \text{error} (\Delta)_{\text{total}} = \pm 0.03 .
\end{equation}

\bibliographystyle{nb}

\begin{thebibliography}{10}
\ifx\href\asklfhas\newcommand{\href}[2]{#2}\fi
\raggedright
\small
\parskip 0pt

\bibitem{Pilch:2000ue}
K.~Pilch and N.~P.~Warner,
\textit{``{N=2 supersymmetric RG flows and the IIB dilaton}''},
\textsf{Nucl.Phys.~B594,~209~(2001)},
\href{http://arXiv.org/abs/hep-th/0004063}{\texttt{hep-th/0004063}}.
%
\bibitem{Buchel:2000cn}
A.~Buchel, A.~W.~Peet and J.~Polchinski,
\textit{``{Gauge dual and noncommutative extension of an N=2 supergravity
  solution}''},
\textsf{Phys.Rev.~D63,~044009~(2001)},
\href{http://arXiv.org/abs/hep-th/0008076}{\texttt{hep-th/0008076}}.
%
\bibitem{Pestun:2007rz}
V.~Pestun,
\textit{``{Localization of gauge theory on a four-sphere and supersymmetric
  Wilson loops}''},
\textsf{Commun.Math.Phys.~313,~71~(2012)},
\href{http://arXiv.org/abs/0712.2824}{\texttt{0712.2824}}.
%
\bibitem{Buchel:2013id}
A.~Buchel, J.~G.~Russo and K.~Zarembo,
\textit{``{Rigorous Test of Non-conformal Holography: Wilson Loops in N=2*
  Theory}''},
\textsf{JHEP~1303,~062~(2013)},
\href{http://arXiv.org/abs/1301.1597}{\texttt{1301.1597}}.
%
\bibitem{Bobev:2013cja}
N.~Bobev, H.~Elvang, D.~Z.~Freedman and S.~S.~Pufu,
\textit{``{Holography for $N = 2^*$ on $S^4$}''},
\textsf{JHEP~1407,~001~(2014)},
\href{http://arXiv.org/abs/1311.1508}{\texttt{1311.1508}}.
%
\bibitem{Chen:2014vka}
X.~Chen-Lin, J.~Gordon and K.~Zarembo,
\textit{``{$ \mathcal{N}={2}^*$ super-Yang-Mills theory at strong coupling}''},
\textsf{JHEP~1411,~057~(2014)},
\href{http://arXiv.org/abs/1408.6040}{\texttt{1408.6040}}.
%
\bibitem{Zarembo:2014ooa}
K.~Zarembo,
\textit{``{Strong-Coupling Phases of Planar N=2* Super-Yang-Mills Theory}''},
\textsf{Theor.Math.Phys.~181,~1522~(2014)},
\href{http://arXiv.org/abs/1410.6114}{\texttt{1410.6114}}.
%
\bibitem{Russo:2013qaa}
J.~G.~Russo and K.~Zarembo,
\textit{``{Evidence for Large-N Phase Transitions in N=2* Theory}''},
\textsf{JHEP~1304,~065~(2013)},
\href{http://arXiv.org/abs/1302.6968}{\texttt{1302.6968}}.
%
\bibitem{Russo:2013kea}
J.~Russo and K.~Zarembo,
\textit{``{Massive N=2 Gauge Theories at Large N}''},
\textsf{JHEP~1311,~130~(2013)},
\href{http://arXiv.org/abs/1309.1004}{\texttt{1309.1004}}.
%
\bibitem{Maldacena:1998im}
J.~M.~Maldacena,
\textit{``{Wilson loops in large N field theories}''},
\textsf{Phys.~Rev.~Lett.~80,~4859~(1998)},
\href{http://arXiv.org/abs/hep-th/9803002}{\texttt{hep-th/9803002}}.
%
\bibitem{Pilch:2003jg}
K.~Pilch and N.~P.~Warner,
\textit{``{Generalizing the N=2 supersymmetric RG flow solution of IIB
  supergravity}''},
\textsf{Nucl.Phys.~B675,~99~(2003)},
\href{http://arXiv.org/abs/hep-th/0306098}{\texttt{hep-th/0306098}}.
%
\bibitem{Fradkin:1984pq}
E.~S.~Fradkin and A.~A.~Tseytlin,
\textit{``{Effective Field Theory from Quantized Strings}''},
\textsf{Phys.~Lett.~B158,~316~(1985)}.
%
\bibitem{Fradkin:1985ys}
E.~S.~Fradkin and A.~A.~Tseytlin,
\textit{``{Quantum String Theory Effective Action}''},
\textsf{Nucl.~Phys.~B261,~1~(1985)}.
%
\bibitem{Luscher:2002qv}
M.~Luscher and P.~Weisz,
\textit{``{Quark confinement and the bosonic string}''},
\textsf{JHEP~0207,~049~(2002)},
\href{http://arXiv.org/abs/hep-lat/0207003}{\texttt{hep-lat/0207003}}.
%
\bibitem{Luscher:1980fr}
M.~Luscher, K.~Symanzik and P.~Weisz,
\textit{``{Anomalies of the Free Loop Wave Equation in the WKB
  Approximation}''},
\textsf{Nucl.~Phys.~B173,~365~(1980)}.
%
\bibitem{Luscher:1980ac}
M.~Luscher,
\textit{``{Symmetry Breaking Aspects of the Roughening Transition in Gauge
  Theories}''},
\textsf{Nucl.~Phys.~B180,~317~(1981)}.
%
\bibitem{Alvarez:1981kc}
O.~Alvarez,
\textit{``{The Static Potential in String Models}''},
\textsf{Phys.~Rev.~D24,~440~(1981)}.
%
\bibitem{Arvis:1983fp}
J.~F.~Arvis,
\textit{``{The Exact $q \bar{q}$ Potential in Nambu String Theory}''},
\textsf{Phys.~Lett.~B127,~106~(1983)}.
%
\bibitem{Olesen:1985pv}
P.~Olesen,
\textit{``{Strings and {QCD}}''},
\textsf{Phys.~Lett.~B160,~144~(1985)}.
%
\bibitem{Greensite:1999jw}
J.~Greensite and P.~Olesen,
\textit{``{World sheet fluctuations and the heavy quark potential in the AdS /
  CFT approach}''},
\textsf{JHEP~9904,~001~(1999)},
\href{http://arXiv.org/abs/hep-th/9901057}{\texttt{hep-th/9901057}}.
%
\bibitem{Forste:1999qn}
S.~Forste, D.~Ghoshal and S.~Theisen,
\textit{``{Stringy corrections to the Wilson loop in N=4 superYang-Mills
  theory}''},
\textsf{JHEP~9908,~013~(1999)},
\href{http://arXiv.org/abs/hep-th/9903042}{\texttt{hep-th/9903042}}.
%
\bibitem{Kinar:1999xu}
Y.~Kinar, E.~Schreiber, J.~Sonnenschein and N.~Weiss,
\textit{``{Quantum fluctuations of Wilson loops from string models}''},
\textsf{Nucl.~Phys.~B583,~76~(2000)},
\href{http://arXiv.org/abs/hep-th/9911123}{\texttt{hep-th/9911123}}.
%
\bibitem{Drukker:2000ep}
N.~Drukker, D.~J.~Gross and A.~A.~Tseytlin,
\textit{``{Green-Schwarz string in $AdS_5\times S^5$: Semiclassical partition
  function}''},
\textsf{JHEP~0004,~021~(2000)},
\href{http://arXiv.org/abs/hep-th/0001204}{\texttt{hep-th/0001204}}.
%
\bibitem{Chu:2009qt}
S.-x.~Chu, D.~Hou and H.-c.~Ren,
\textit{``{The Subleading Term of the Strong Coupling Expansion of the
  Heavy-Quark Potential in a N=4 Super Yang-Mills Vacuum}''},
\textsf{JHEP~0908,~004~(2009)},
\href{http://arXiv.org/abs/0905.1874}{\texttt{0905.1874}}.
%
\bibitem{Forini:2010ek}
V.~Forini,
\textit{``{Quark-antiquark potential in AdS at one loop}''},
\textsf{JHEP~1011,~079~(2010)},
\href{http://arXiv.org/abs/1009.3939}{\texttt{1009.3939}}.
%
\bibitem{Gromov:2016rrp}
N.~Gromov and F.~Levkovich-Maslyuk,
\textit{``{Quark-anti-quark potential in $ \mathcal{N} =$ 4 SYM}''},
\textsf{JHEP~1612,~122~(2016)},
\href{http://arXiv.org/abs/1601.05679}{\texttt{1601.05679}}.
%
\bibitem{Drukker:2012de}
N.~Drukker,
\textit{``{Integrable Wilson loops}''},
\textsf{JHEP~1310,~135~(2013)},
\href{http://arXiv.org/abs/1203.1617}{\texttt{1203.1617}}.
%
\bibitem{Correa:2012hh}
D.~Correa, J.~Maldacena and A.~Sever,
\textit{``{The quark anti-quark potential and the cusp anomalous dimension from
  a TBA equation}''},
\textsf{JHEP~1208,~134~(2012)},
\href{http://arXiv.org/abs/1203.1913}{\texttt{1203.1913}}.
%
\bibitem{Cvetic:1999zs}
M.~Cvetic, H.~Lu, C.~N.~Pope and K.~S.~Stelle,
\textit{``{T-Duality in the Green-Schwarz Formalism, and the Massless/Massive
  IIA Duality Map}''},
\textsf{Nucl.~Phys.~B573,~149~(2000)},
\href{http://arXiv.org/abs/hep-th/9907202}{\texttt{hep-th/9907202}}.
%
\bibitem{tHooft:1976snw}
G.~'t~Hooft,
\textit{``{Computation of the Quantum Effects Due to a Four-Dimensional
  Pseudoparticle}''},
\textsf{Phys.~Rev.~D14,~3432~(1976)}.
%
\bibitem{Kruczenski:2008zk}
M.~Kruczenski and A.~Tirziu,
\textit{``{Matching the circular Wilson loop with dual open string solution at
  1-loop in strong coupling}''},
\textsf{JHEP~0805,~064~(2008)},
\href{http://arXiv.org/abs/0803.0315}{\texttt{0803.0315}}.
%
\bibitem{Dashen:1974cj}
R.~F.~Dashen, B.~Hasslacher and A.~Neveu,
\textit{``{Nonperturbative Methods and Extended Hadron Models in Field Theory.
  2. Two-Dimensional Models and Extended Hadrons}''},
\textsf{Phys.~Rev.~D10,~4130~(1974)}.
%
\bibitem{Dimov:2003bh}
H.~Dimov, V.~G.~Filev, R.~Rashkov and K.~Viswanathan,
\textit{``{Semiclassical quantization of rotating strings in Pilch-Warner
  geometry}''},
\textsf{Phys.Rev.~D68,~066010~(2003)},
\href{http://arXiv.org/abs/hep-th/0304035}{\texttt{hep-th/0304035}}.
%
\bibitem{Young:2014jma}
D.~Young and K.~Zarembo,
\textit{``{Holographic Dual of the Eguchi-Kawai Mechanism}''},
\textsf{JHEP~1406,~030~(2014)},
\href{http://arXiv.org/abs/1404.0225}{\texttt{1404.0225}}.
%
\bibitem{Bigazzi:2001aj}
F.~Bigazzi, A.~Cotrone and A.~Zaffaroni,
\textit{``{N=2 gauge theories from wrapped five-branes}''},
\textsf{Phys.Lett.~B519,~269~(2001)},
\href{http://arXiv.org/abs/hep-th/0106160}{\texttt{hep-th/0106160}}.
%
\bibitem{Bigazzi:2003ui}
F.~Bigazzi, A.~L.~Cotrone, M.~Petrini and A.~Zaffaroni,
\textit{``{Supergravity duals of supersymmetric four-dimensional gauge
  theories}''},
\textsf{Riv.~Nuovo~Cim.~25N12,~1~(2002)},
\href{http://arXiv.org/abs/hep-th/0303191}{\texttt{hep-th/0303191}}.
%
\bibitem{Bigazzi:2013xia}
F.~Bigazzi, A.~L.~Cotrone, L.~Griguolo and D.~Seminara,
\textit{``{A novel cross-check of localization and non conformal
  holography}''},
\textsf{JHEP~1403,~072~(2014)},
\href{http://arXiv.org/abs/1312.4561}{\texttt{1312.4561}}.
%
\bibitem{Polchinski:2000uf}
J.~Polchinski and M.~J.~Strassler,
\textit{``{The String dual of a confining four-dimensional gauge theory}''},
\href{http://arXiv.org/abs/hep-th/0003136}{\texttt{hep-th/0003136}}.
%
\bibitem{Bobev:2016nua}
N.~Bobev, H.~Elvang, U.~Kol, T.~Olson and S.~S.~Pufu,
\textit{``{Holography for $ \mathcal{N} $ = 1$^{∗}$ on S$^{4}$}''},
\textsf{JHEP~1610,~095~(2016)},
\href{http://arXiv.org/abs/1605.00656}{\texttt{1605.00656}}.
%
\bibitem{Metsaev:1998it}
R.~R.~Metsaev and A.~A.~Tseytlin,
\textit{``Type IIB superstring action in $AdS_5 \times S^5$ background''},
\textsf{Nucl.~Phys.~B533,~109~(1998)},
\href{http://arXiv.org/abs/hep-th/9805028}{\texttt{hep-th/9805028}}.
%
\bibitem{Buchbinder:2014nia}
E.~Buchbinder and A.~Tseytlin,
\textit{``{The 1/N correction in the D3-brane description of circular Wilson
  loop at strong coupling}''},
\textsf{Phys.Rev.~D89,~126008~(2014)},
\href{http://arXiv.org/abs/1404.4952}{\texttt{1404.4952}}.
%
\end{thebibliography}

\end{document}